\documentclass[conference, letterpaper]{IEEEtran}
\makeatletter
\def\ps@IEEEtitlepagestyle{%
  \def\@oddfoot{\mycopyrightnotice}%
}
\def\mycopyrightnotice{%
  \begin{minipage}{\textwidth}
  \centering \scriptsize
  Copyright~\copyright~2023 IEEE. Personal use of this material is permitted. Permission from IEEE must be obtained for all other uses, in any current or future media, including\\reprinting/republishing this material for advertising or promotional purposes, creating new collective works, for resale or redistribution to servers or lists, or reuse of any copyrighted component of this work in other works by sending a request to pubs-permissions@ieee.org. 
  
  DOI: 10.1109/CLUSTER52292.2023.00023
  \end{minipage}
}
\makeatother

\usepackage{graphicx}
\usepackage{caption}
\usepackage{subcaption}
\usepackage{svg}
\graphicspath{{./img/}}
\usepackage{array}
\newcolumntype{P}[1]{>{\centering\arraybackslash}p{#1}}
\newcolumntype{M}[1]{>{\centering\arraybackslash}m{#1}}
\usepackage{url}

\usepackage{flushend}
\usepackage{amsfonts}

\begin{document}

\title{Hierarchical Resource Partitioning on Modern GPUs: A Reinforcement Learning Approach
}

\author{
\IEEEauthorblockN{
Urvij Saroliya, 
Eishi Arima, 
Dai Liu, and 
Martin Schulz
}
\IEEEauthorblockA{		
Technical University of Munich, Garching, Germany
\\\{urvij.saroliya, eishi.arima, dai.liu, martin.w.j.schulz\}@tum.de
}
}

\maketitle

\begin{abstract}
GPU-based heterogeneous architectures are now commonly used in HPC clusters. 
Due to their architectural simplicity specialized for data-level parallelism, GPUs can offer much higher computational throughput and memory bandwidth than CPUs in the same generation do. 
However, as the available resources in GPUs have increased exponentially over the past decades, it has become increasingly difficult for a single program to fully utilize them. 
As a consequence, the industry has started supporting several resource partitioning features in order to improve the resource utilization by co-scheduling multiple programs on the same GPU die at the same time. 

Driven by the technological trend, this paper focuses on hierarchical resource partitioning on modern GPUs, and as an example, we utilize a combination of two different features available on recent NVIDIA GPUs in a hierarchical manner: MPS (Multi-Process Service), a finer-grained logical partitioning; and MIG (Multi-Instance GPU), a coarse-grained physical partitioning. 
We propose a method for comprehensively co-optimizing the setup of hierarchical partitioning and the selection of co-scheduling groups from a given set of jobs, based on reinforcement learning using their profiles. 
Our thorough experimental results demonstrate that our approach can successfully set up job concurrency, partitioning, and co-scheduling group selections simultaneously. 
This results in a maximum throughput improvement by a factor of 1.87 compared to the time-sharing scheduling. 
\end{abstract}


\section{Introduction}\label{Introduction}
HPC clusters and supercomputers are becoming increasingly heterogeneous, and as a consequence, 172 out of 500 top-class supercomputers are now GPU-equipped systems (as of Nov 2022)~\cite{top500}. 
This trend has started ever since Dennard scaling ceased over a decade ago~\cite{dennard, dennard2}. 
As  single-thread performance improvements were sustained by Dennard scaling, the industry had to shift towards multi-/many-core processors and heterogeneous architectures focusing on thread-/data-level parallelisms. 
GPUs are specialized hardware to exploit data-level parallelism of applications by spending more transistors on  compute resources and simplifying the control logic considerably compared with those of CPUs. 
By taking advantage of this simplicity, GPUs can offer much higher computational throughput and memory bandwidth than CPUs in the same VLSI technology generation (typically several times higher). 

However, as the available resources in GPUs have increased exponentially over the past decades, it has become increasingly difficult for a single program to fully utilize them. 
The first reason for this is that not all GPU programs have sufficient parallelism to convert the available compute resources inside a GPU into speedup, which is governed by the well-known Amdahl's law~\cite{amdahl}. 
The second reason is the throughput of memory intensive applications is limited by the available memory bandwidth, and thus increasing the compute resources does not contribute to the speedup for them, which is known as the memory-wall problem~\cite{memory-wall}.
The third reason is the compute resources inside a GPU are also becoming heterogeneous with different types of units (e.g., matrix engines, regular FP64 units, integer units, etc.), and depending on their usages, power can also be under utilized and wasted~\cite{gpu-mig-power}. 

As a consequence, the industry has started supporting several resource partitioning features that enable multiple programs to be co-scheduled on the same GPU at the same time with variable resource allocations. 
One example is MPS (Multi-Process Service) that allows multiple programs to share computational resources \textit{logically}, which is supported in recent NVIDIA GPUs~\cite{mps}. 
The MPS feature is a software-based mechanism with several architectural supports that decides the process to SM (Streaming Processor) assignments with arbitrary rates (e.g., 70\%). 
Another example is MIG (Multi-Instance GPU) that \textit{physically} partitions computational and bandwidth resources in a hierarchical manner at the granularity of GPC (Graphics Processing Cluster), which is supported in recent high-end NVIDIA GPUs from the Ampere generation~\cite{mig}. It first partitions a GPU into one or more GIs (GPU Instances), each of which is completely isolated, and then partitions each GI into one or more CIs (Compute Instances) that share the memory resources within the GI but utilize the compute resources mutually exclusively at the granularity of GPC. 
These different partitioning features can be applied at the same time in a hierarchical manner, i.e., the MPS is applicable inside a CI created by the MIG.

This paper explicitly targets hierarchical resource partitioning on modern GPUs, e.g., the hierarchical combination of MIG (coarse-grained physical partitioning) and MPS (fine-grained logical resource allocations), and orchestrates the multi-level partitioning setup and co-scheduling decision making for a given set of jobs. 
To this end, we first analyze the impact of partitioning setup on performance using different workloads 
and demonstrate the potential benefit of the hierarchical partitioning. 
Driven by the observations, we propose our resource management method based on reinforcement learning using job profiles. 
More specifically, we regard the optimization as a classification problem and choose an optimal set of partitioning and co-scheduling groups for a given set of jobs. 
We train the classifier composed of a deep Q network using reinforcement learning at offline and apply the optimized agent to the online optimization. 

Specifically, this paper makes the following major contributions: 
\begin{itemize}
\item As far as we know, this work is the first to apply reinforcement learning to simultaneously optimizing job co-scheduling and hierarchical resource partitioning on platforms with multiple different partitioning features (MPS and MIG) available on recent commercial GPUs. 
\item We quantitatively analyze the impact of the resource partitioning setup on GPUs throughput, while comparing different partitioning setups.  
\item We then define the co-scheduling and resource partitioning process as an optimization problem in a concrete mathematical form, which enables us to regard the optimization as a classification problem and introduce a reinforcement learning-based solution. 
\item We demonstrate that our approach can successfully set up the partitioning and co-scheduling group selections simultaneously through our thorough evaluations, and discuss how it is extensible to the entire cluster scale. 
\end{itemize}


\section{Related Work}\label{Literature}
Since multi-/many-core architectures and CPU-GPU heterogeneous architectures became common in servers and HPC clusters, a variety of co-scheduling and resource partitioning techniques have been proposed. 
However, as far as we know, \textit{this is the first work to apply a reinforcement learning approach to co-optimize the hierarchical resource partitioning and co-scheduling job selections for modern GPUs. }

\noindent\textbf{Literature of Co-scheduling and Resource Partitioning: }
Since multi-core processors appeared on the market, several researchers have proposed co-scheduling mechanisms while focusing on multi-programmed but single-threaded workloads. 
Y. Jiang et al. studied the theoretical aspects of co-scheduling and provided an optimal solution~\cite{cos-pact0}. 
Then, K. Tai et al. extended this theoretical work to take the execution time lengths into account~\cite{cos-cf}. 
S. Zhuravlev et al. focused on the shared resource contention in a processor and proposed an interference-aware co-scheduling method~\cite{cos-asplos}. 
J. Feliu et al. proposed a scheduling policy that explicitly considers the contentions on the underlying shared cache hierarchy~\cite{cos-cache}. 
M. Banikazemi et al. designed and implemented a user-level meta co-scheduler and demonstrated the effectiveness~\cite{cos-sc}. 

Other researchers extended the ideas and proposed several co-scheduling techniques for multi-threaded programs. 
M. Bhadauria et al. explored the feasibility of space-shared scheduling using a greedy-based co-run job selection and resource allocation policy~\cite{cos-ics}. Then, H. Sasaki et al. proposed a scalability-based resource allocation approach for a given multi-programmed and multi-threaded workload~\cite{cos-pact}. J. Breitbart et al. created a resource monitoring tool useful for co-scheduling HPC applications~\cite{cos-bw} and provided a memory-intensity-aware co-scheduling policy~\cite{cos-bw2}. 
Since the industry started supporting several QoS control features, some researchers combined the above concepts with cache partitioning~\cite{cos-cp,cos-cp2}, bandwidth partitioning~\cite{cos-bp,cos-bp2} or the combination of them~\cite{cos-cbp,rl-memctrl}. 
Q. Zhu et al. rather targeted CPU-GPU heterogeneous processors and proposed a co-scheduling approach suitable for them~\cite{cos-hetero}.

\noindent\textbf{Applying Co-scheduling and Resource Partitioning to GPUs: }
S. Pai et al. first pointed out the waste of resources within a GPU when running a CUDA kernel and explored the feasibility of GPU multi-processing using their elastic kernel implementation~\cite{elastic-kernel}. 
I. Tanasic et al. proposed a microarchitectural mechanism to enable multi-processing on GPUs, which does not require any kernel modifications~\cite{preemptive-gpu}. 
Following these seminal studies, the MPS feature has been already supported in commercial Nvidia GPUs~\cite{mps}.

Several studies focused on software mechanisms to improve the efficiency of multi-processing on GPUs. 
T. Allen et al. proposed a framework called Slate that optimizes the combination of co-located processes and dynamically adjusts the scales of them~\cite{slate}. 
smCompactor is a similar framework to Slate, which aims at maximizing the resource utilization~\cite{smcompactor}. 
C. Reano et al. proposed a safe co-scheduling mechanism that takes memory footprints into account when processes are co-scheduled in a time sharing manner~\cite{time-sharing}. 
Other studies rather focused on hardware mechanisms to improve the efficiency of the concurrency controlling features~\cite{hw-space-sharing, hw-space-sharing2, hw-space-sharing3}.
Since the industry has started supporting the physical resource partitioning called MIG~\cite{mig}, few studies targeted the MIG-based partitioning and proposed several optimization mechanisms~\cite{mig-miso, gpu-mig-power, cos-hetero2}. 
The closest work to ours is~\cite{cos-hetero2}, which covers co-scheduling decision making and resource partitioning, however it does not manage the \textit{hierarchical partitioning} and works only when co-locating \textit{two} programs. 



\noindent\textbf{System Optimizations with Reinforcement Learning:} 
Since reinforcement learning is a powerful tool to optimize software or hardware knobs, it has been widely used also for a variety of system optimizations. 
Although these techniques are promising or already widely used, they target fundamentally different problems from ours. 
E. Ipek et al. proposed a reinforcement learning-based memory controller design that optimizes the scheduling policy on the fly~\cite{rl-memctrl}. 
Following this seminal work, there have been a variety of software/hardware optimizations using reinforcement learning.  
Yoo et al. applied reinforcement learning to determine several parameters in a QLC SSD such as the SLC cache size and the hot/cold separation threshold~\cite{rl-slc}. 
D. Zhang et al. invented RLScheduler that automatically configures the priority function used for batch scheduling in HPC systems based on reinforcement learning~\cite{rl-sched}. 
R. Chen et al. utilized reinforcement learning to co-optimize the cache and bandwidth allocations for multi-programmed server workloads~\cite{rl-partition}. 
Y. Wang et al. proposed a power management technique for multi-core processors based on reinforcement learning~\cite{rl-power}. 
P. Zhang et al. applied an reinforcement learning approach to an ensemble controller that dynamically selects the best prefetch policy from multiple different prefetchers~\cite{rl-prefetchers}. 
G. Singh et al. targeted hybrid storage systems and proposed an adaptive and extensible data placement using their online reinforcement learning approach~\cite{rl-hss}. 
\section{Observations}\label{Motivation}
In this section, we observe the effectiveness and performance impact of hierarchical partitioning using the combination of MIG and MPS features as an example. 
In Section~\ref{background}, we introduce the summary of these two partitioning features and how they are configured in a hierarchical fashion. 
In Section~\ref{observation}, we demonstrate the impact of the partitioning setup on performance and analyze it based on the characteristics of co-located applications.

\begin{figure}[t]
\begin{center}
\includegraphics[width=\linewidth]{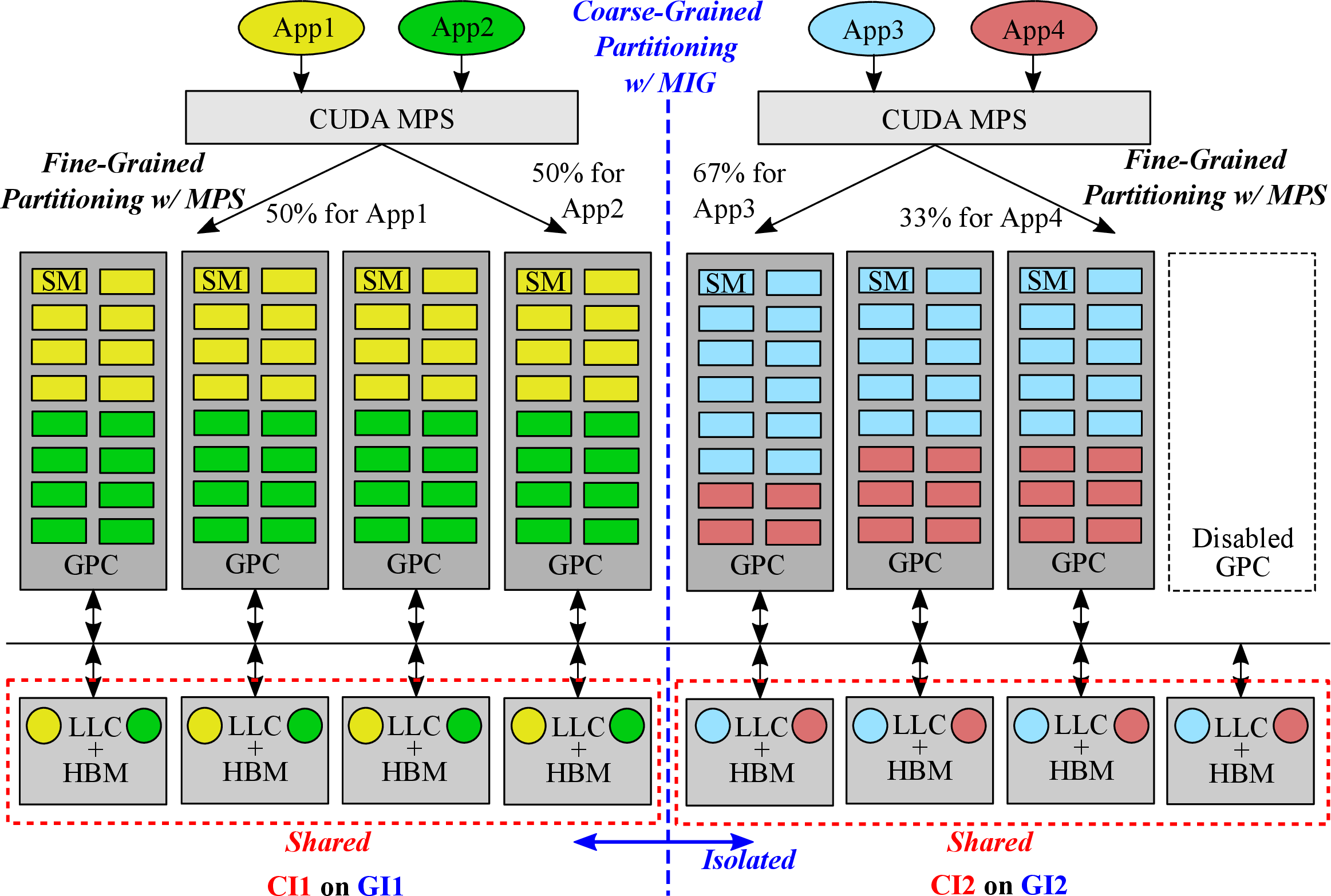}
\caption{A modern GPU architecture and a hierarchical partitioning on it}\label{mig-mps-partitioning}
\end{center}
\end{figure}

\subsection{Hierarchical Partitioning on Modern GPUs}\label{background}

Figure~\ref{mig-mps-partitioning} illustrates a modern GPU architecture and our target hierarchical partitioning, e.g., NVIDIA Ampere architecture~\cite{a100} and the combination of MIG~\cite{mig} and MPS~\cite{mps} features. 
In order to enable massive parallelism, modern GPUs are structured in a hierarchical manner. 
In the NVIDIA Ampere architecture, as an example, one GPU consists of multiple GPCs (Graphics Processing Clusters), and each GPC is composed of multiple SMs (Streaming Processors). 
On one hand, one SM has its own private resources including a local instruction/data cache, a warp scheduler, a dispatcher, a register file, and many functional units (e.g., FPUs, ALUs, LSUs, a matrix engine, etc.). 
On the other hand, there are shared resources such as LLCs (Last Level Caches) and device memory blocks (HBM stacks) reachable by any GPCs by default. 

A GPU can be partitioned in the following way. First, with the MIG feature, a GPU is divided into one or more GIs (GPU Instances) at the granularity of GPC, and then one or more CIs (Compute Instances) are launched on each GI while occupying  GPCs within the GI in a mutually exclusive manner. 
Then, the user selects one of the CIs and run a program on it. 
One GI owns the same number of LLC/HBM blocks as that of GPCs, and they become private and isolated resources accessible only by the CIs launched on the GI. 
As the MIG feature is a coarse-grained physical resource partitioning, it is not flexible, and there are several restrictions: (1) one GPC needs to be disabled when turning on the feature; (2) it is configurable only when no program is running; and (3) the partitioning choices are limited only to 19 variants in the current implementation --- for instance, dividing 7GPCs into (a) 2GPCs and 5GPCs or (b) 1GPCs and 6GPCs are not supported~\cite{mig}.

Second, each CI (or the entire GPU if the MIG is not applied) can be partitioned further at the granularity of SM with the MPS feature. 
On one hand, the MPS partitioning is more flexible and finer-grained than that of the MIG feature including both the GI- and CI-level partitioning. 
On the other hand, it does not offer any knobs to control the quality of service (e.g., shared resource partitioning). 
Therefore, \textit{the MIG feature should be used for setting up the shared memory resource partitioning/isolation to mitigate the interference impact, however the MPS is useful to flexibly assign the compute resources to balance the performance of all the co-located programs (better than the CI-level partitioning).}

\begin{figure}[t]
\begin{center}
\includegraphics[width=\linewidth]{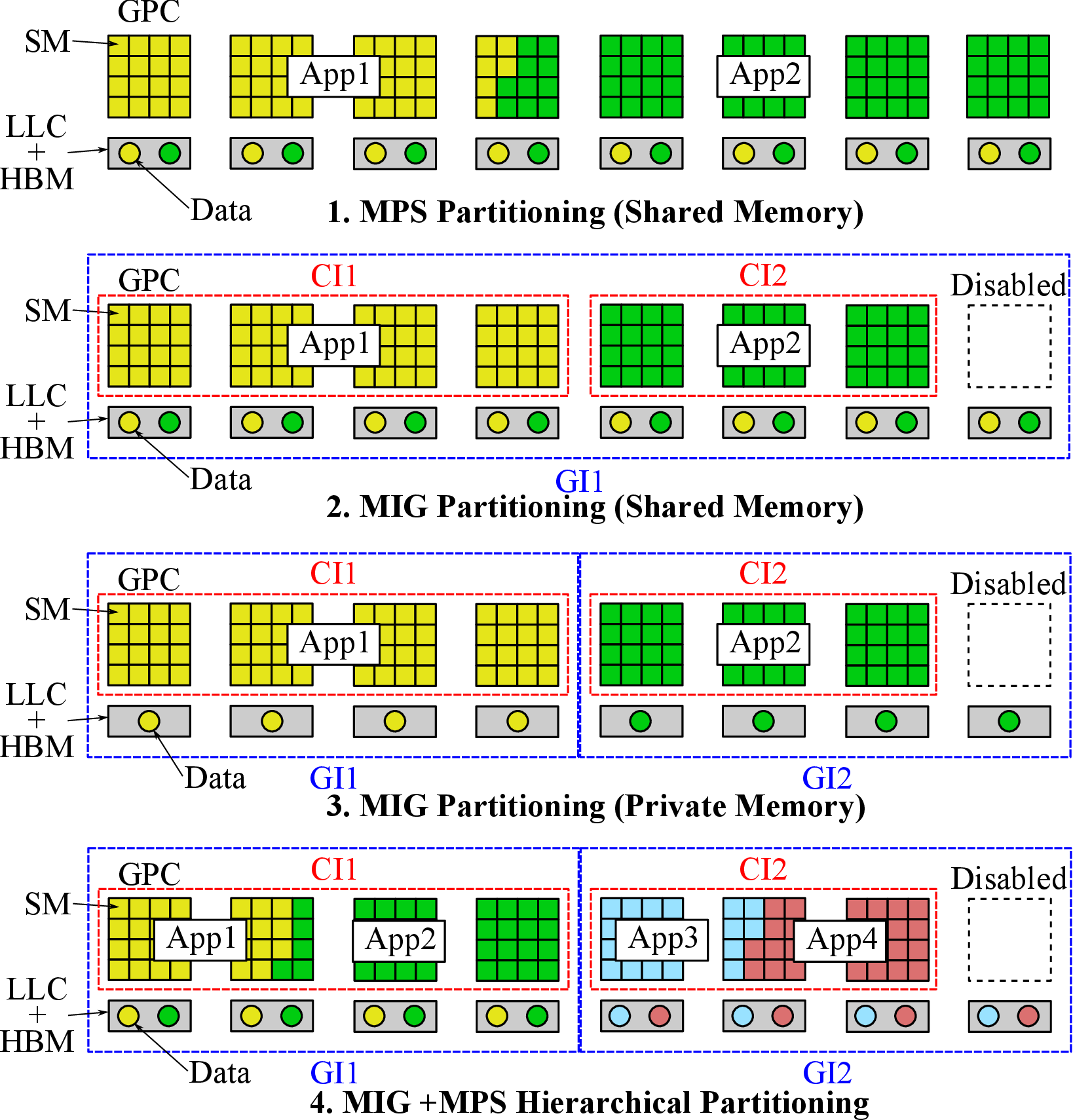}
\caption{Partitioning variations with MIG and MPS}\label{partitioning-variants}
\end{center}
\end{figure}

\begin{figure*}[t]
    \includegraphics[width=0.246\linewidth]{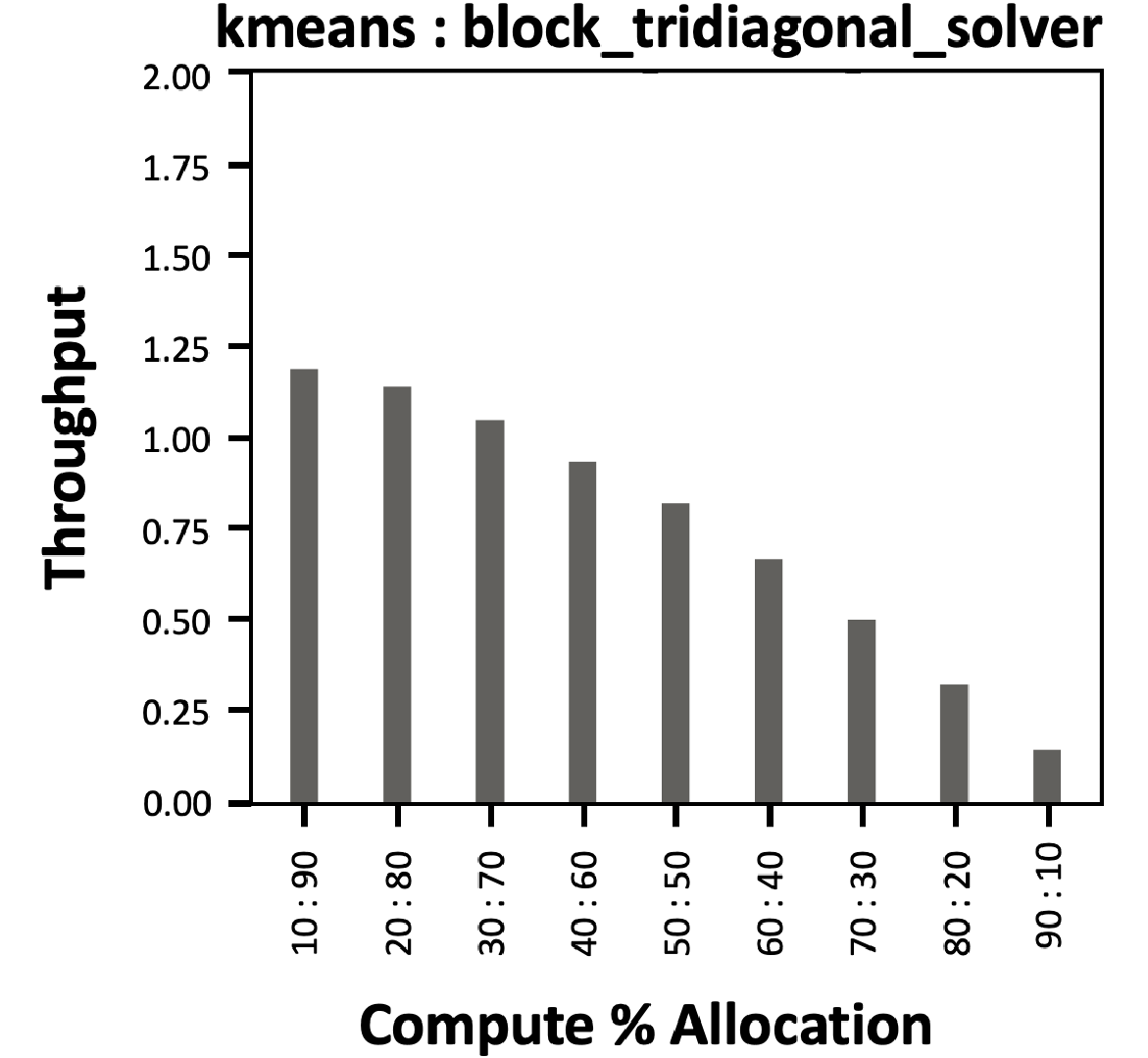}
    \includegraphics[width=0.246\linewidth]{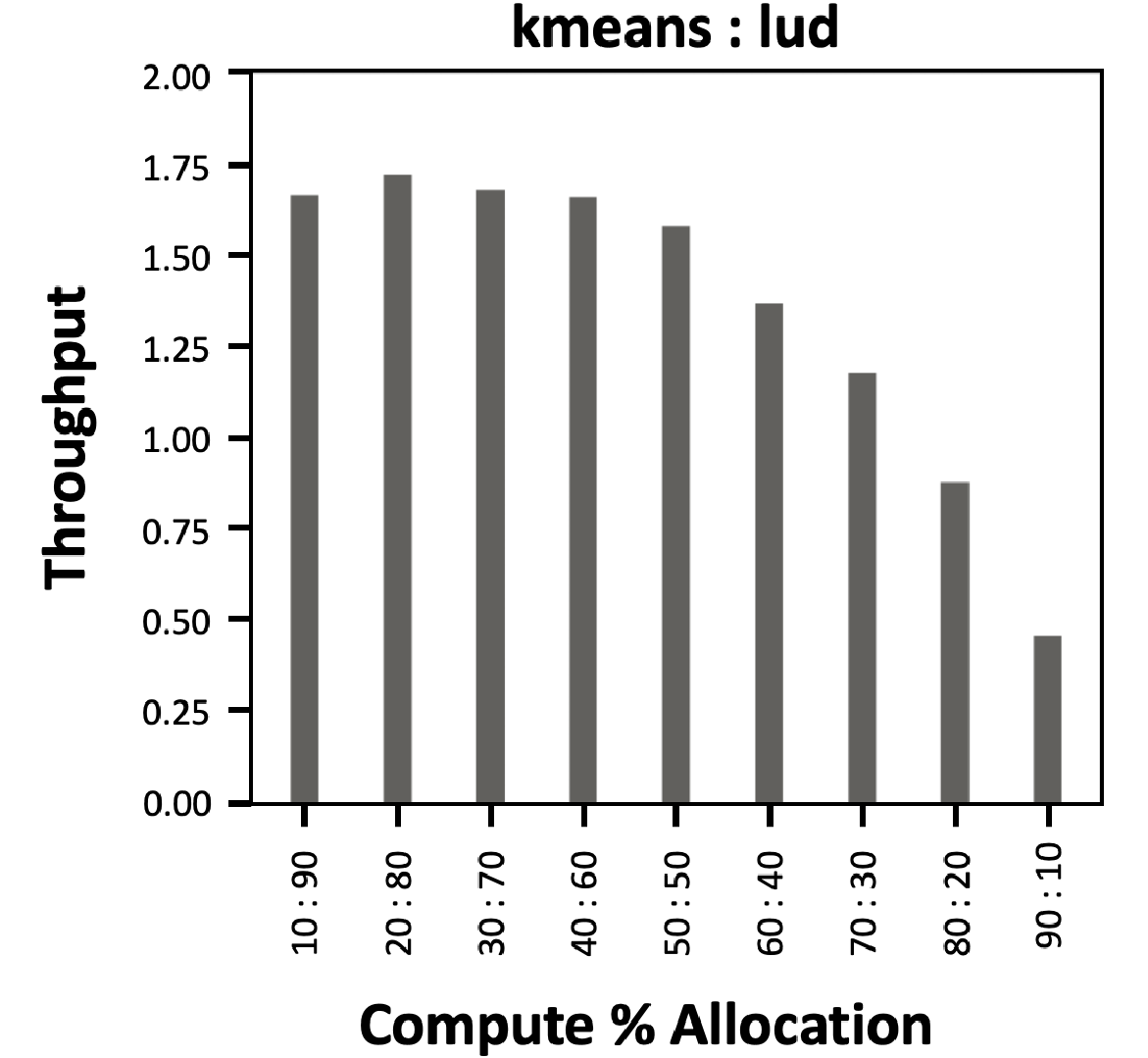}
    \includegraphics[width=0.246\linewidth]{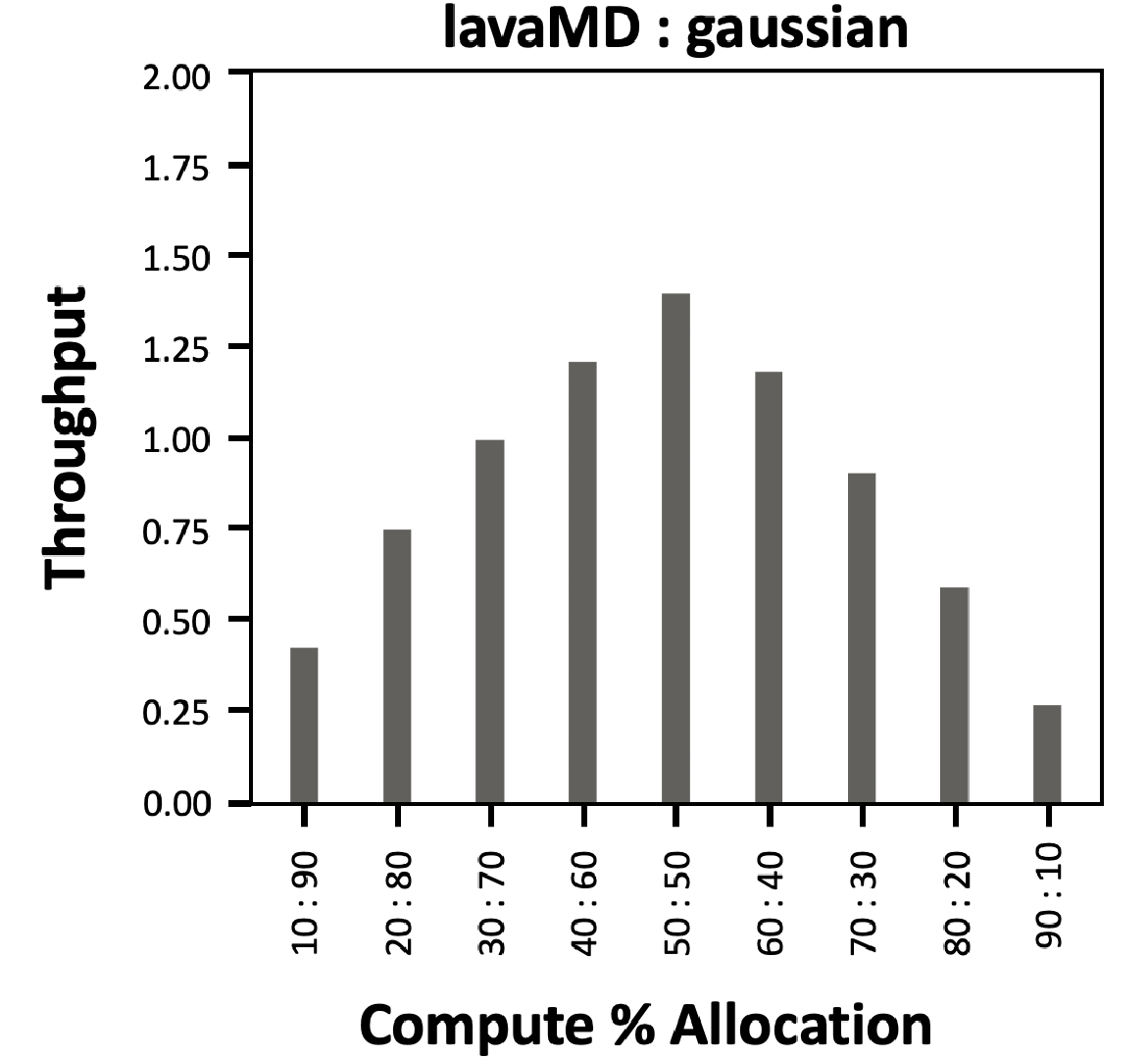}
    \includegraphics[width=0.246\linewidth]{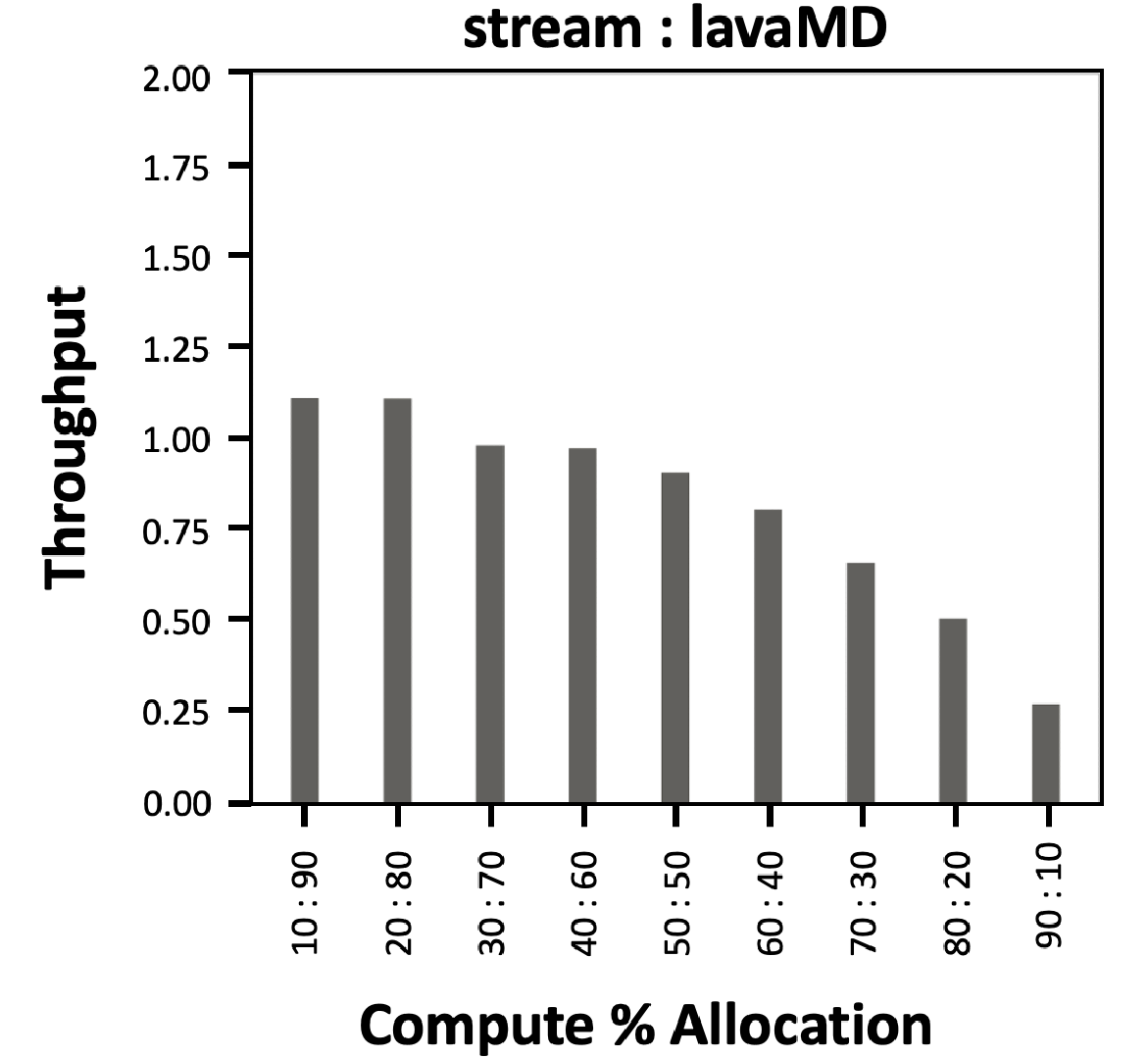}    
    \caption{Co-scheduling Throughput as a Function of Compute Resource Allocations (MPS Partitioning)}
    \label{fig:mps-only}
\end{figure*}

\begin{figure}[t]
\vspace{-10pt}
\begin{center}
\includegraphics[height=4cm]{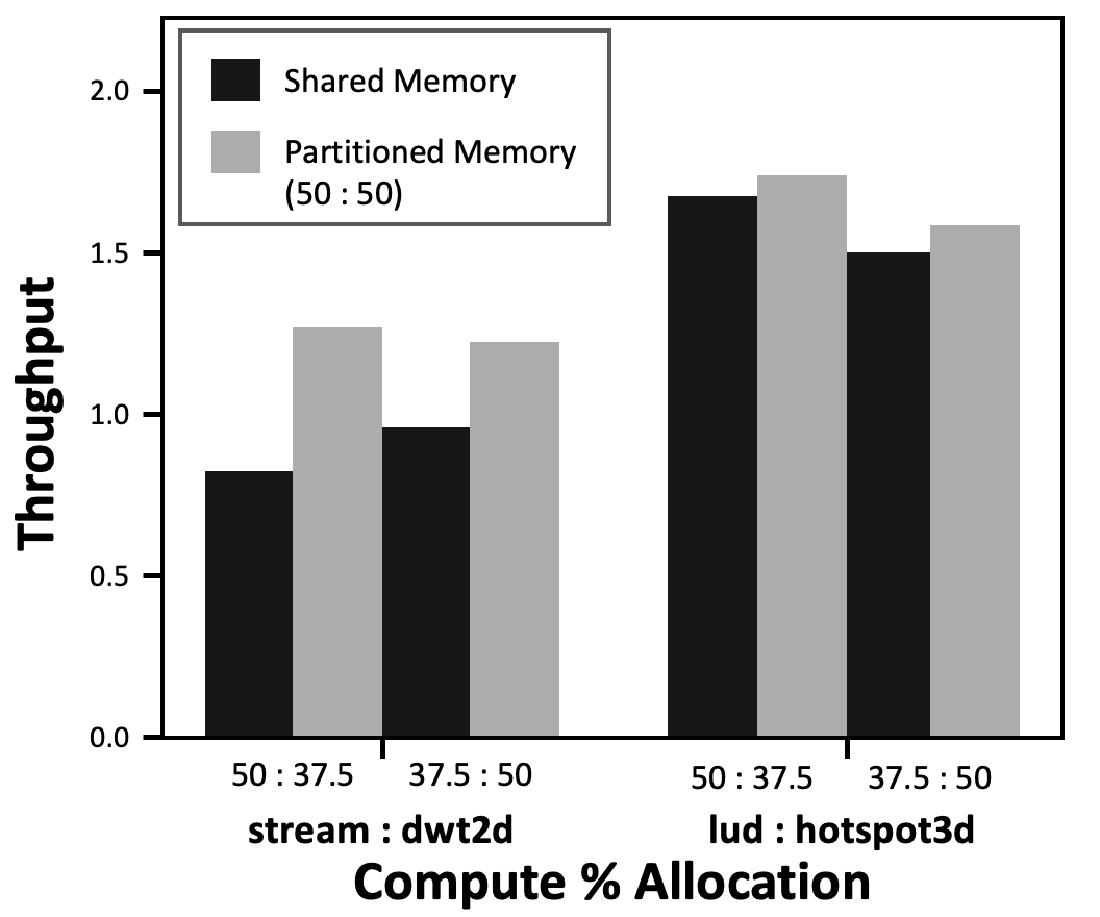}
\caption{Performance Benefit of Bandwidth Partitioning}\label{mig-comparison}
\end{center}
\end{figure}

The combination of these two features can offer multiple different partitioning variations, as shown in Figure~\ref{partitioning-variants}. 
\textbf{The first two options} do not partition the memory resources, but share memory across all co-located applications. 
These options are useful when the co-located applications require \textit{complementary resources}, i.e., one is a \textit{compute-bound} application that does not fully utilize the available memory bandwidth, and the other one is rather \textit{memory bound}, for which only a small subset of the available compute resources is enough. 
The MPS-only option has more advantages than the MIG-only shared memory option: (1) the MPS can set the compute resource allocations in a more flexible and fine-grained manner; and (2) the MIG needs to turn off 1 out of 8 GPCs, while the MPS can utilize all available 8 GPCs (for an A100 GPUs~\cite{a100}). 

\textbf{The third option} in the figure is useful to mitigate \textit{shared resource conflicts} among co-located applications. 
This interference-free option is effective in particular for \textit{not well scalable} applications, as the option limits both the compute and bandwidth resources on the GPU at the same time. 
As Amdahl's law suggests~\cite{amdahl}, the scalability is limited by the program's parallelism (or the overhead of parallelization), which is also the case for GPU applications limited by issues such as synchronization overhead or problem size. 
This scalability limit inside a GPU will be even more serious when the compute/bandwidth resources become richer due to further scaling of VLSI technology in the future.

Finally, \textbf{the last option} is the mixture of MIG and MPS as a general form and an intermediate case of all the above options. 
This approach is promising, especially when we execute more programs concurrently on the GPU, and it is suitable for a variety of program mixes. 
We regard the first three options as extreme setups of this hierarchical partitioning approach. 
When we co-locate more than two programs inside a GI, we increase the concurrency in the MPS, while setting the number of CIs to 1, as this allows us to use the full flexibility of the MPS feature.

\subsection{Observational Analysis}\label{observation}

Figure~\ref{fig:mps-only} demonstrates GPU throughput as a function of compute resource allocation to two co-located HPC benchmark programs across different program mixes. 
In this evaluation, we utilize the MPS-based partitioning as illustrated in the first option of \figurename{ \ref{partitioning-variants}}. 
The X-axis represents the ratios of compute resource allocation to the co-scheduled programs shown at the legend, while the Y-axis indicates the relative throughput normalized to that of a time-sharing scheduling, i.e., executing these two programs one by one without sharing the resources but with fully allocating the entire GPU resources. 
As illustrated in \figurename{ \ref{fig:mps-only}}, the optimal allocation of compute resources to the co-located programs depend highly on the given programs and their characteristics. 
As we can observe in the third case, a balanced allocation achieves the best performance, while for the others, a skewed allocation has advantage over a balanced one with a unique optimal allocation point. 
With such varying optimal allocations for different program mixes, 
we conclude that \textit{compute resource partitioning features need to be fine-grained and flexible so that one can fine-tune the allocation setup, and MPS is more preferable for this purpose. 
}

\begin{figure}[t]
\vspace{-10pt}
\begin{center}
\includegraphics[height=4cm]{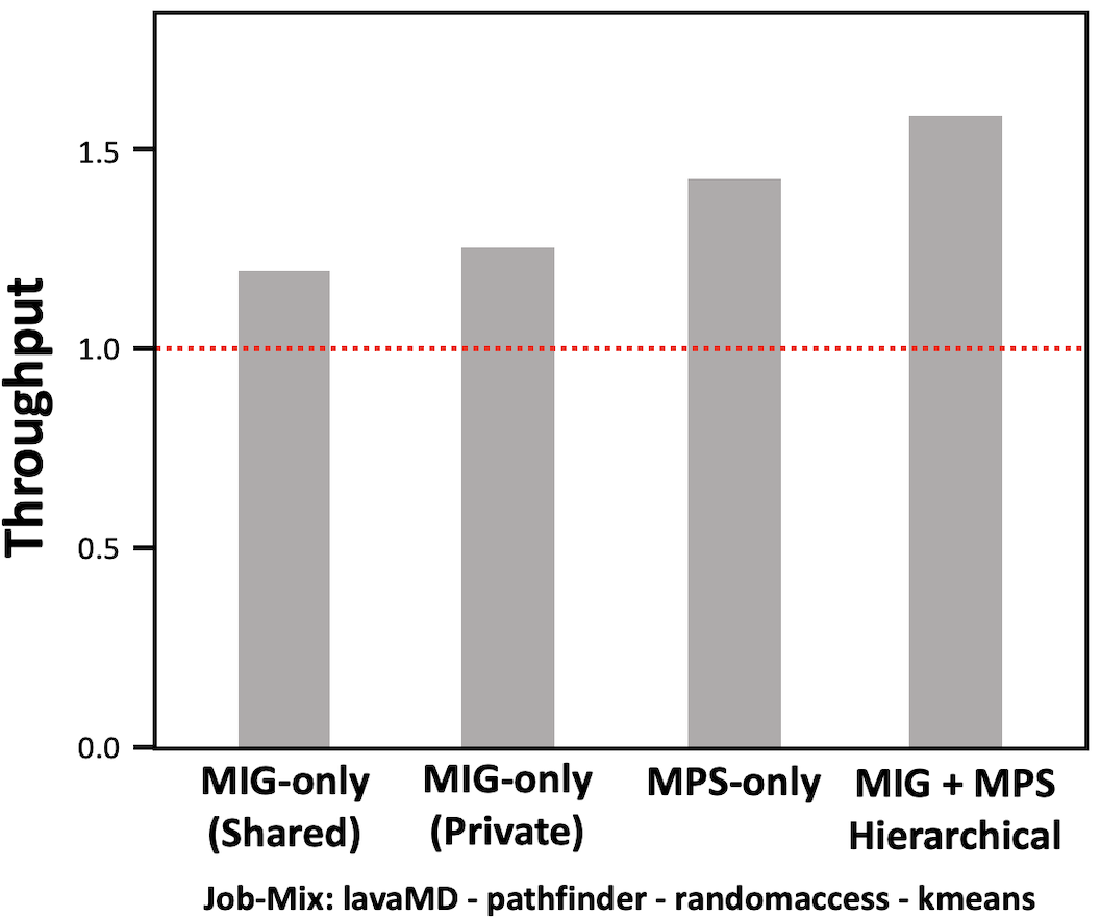}
\caption{Performance Comparison for Different Partitioning Variants Introduced in Section~\ref{background}}\label{variant-comparison}
\end{center}
\end{figure}



Figure~\ref{mig-comparison} presents the impact of memory bandwidth resource partitioning while using the two different MIG options (shared or private) introduced in Figure~\ref{partitioning-variants}. 
The X-axis lists two different job mixes with two different compute resource allocation rates as well as two different memory options (shared or partitioned), while the Y-axis shows the relative throughput normalized to that of the time-sharing scheduling as mentioned above. 
To assess the impact of memory partitioning on performance, we setup exact the same compute resource allocation for the shared and partitioned options. One GPC is disabled in this evaluation, and thus the total of the compute resource allocation percentages is 87.5\% in each case. 
For these job mixes, we observe considerable speedup by partitioning/isolating memory bandwidth resources by mitigating the interference impact among the co-located programs. 
Therefore, \textit{
depending on the given job mix, it is preferable to partition/isolate the shared memory resources in order to mitigate the interference impact, and only the MIG feature is useful for this purpose. 
}


Finally, Figure~\ref{variant-comparison} compares multiple different partitioning options illustrated in Figure~\ref{partitioning-variants}. 
The horizontal axis lists all the options introduced in Figure~\ref{partitioning-variants}, while the vertical axis indicates the relative throughput normalized to that of the time-sharing scheduling mentioned above. 
The job mix shown in the legend of the figure lists 4 programs to be co-scheduled, and the pairs are selected optimally for each partitioning option.
The best compute resource allocation [\%] is selected to given two co-located programs for the \textit{MPS Only} option. For the \textit{MIG Only} options, each co-located application is assigned to one of the 4GPC or 3GPC CIs, which is selected optimally so that the throughput is maximized. 
The \textit{MIG+MPS Hierarchical} is a mixture of these options. We co-locate all four programs at the same time on the GPU. We first partition it into 4GPCs and 3GPCs with the MIG feature, and then each of the co-located programs is assigned to one of them with optimal compute resource allocations [\%] designated by the MPS feature. 
Note that we search the optimal setups as well as the job pair selections in an exhaustive manner for all the above options. 
As shown in the figure, by combining the two different partitioning features in a hierarchical manner, we observe even more throughput improvement. 



\section{Our Approach}\label{Proposal}
As demonstrated in the previous section, hierarchical resource partitioning, using a combination of MIG and MPS features, is effective to improve the throughput of GPUs. 
However, the partitioning setup needs to be chosen carefully as the best choice highly depends on the characteristics of co-located programs.  
At the same time, the selection of jobs to co-schedule from a given job queue also significantly affects system performance. 
In this paper, we target both the co-scheduling and resource partitioning decisions and co-optimize them simultaneously. 
To this end, we first formulate the decision making as an optimization problem. 
Second, we design a reinforcement learning-based co-scheduling and resource management system to solve the problem, which consists of offline profiling/training and online optimization. 


\begin{figure}
\begin{center}
  \includegraphics[width=0.9\linewidth]{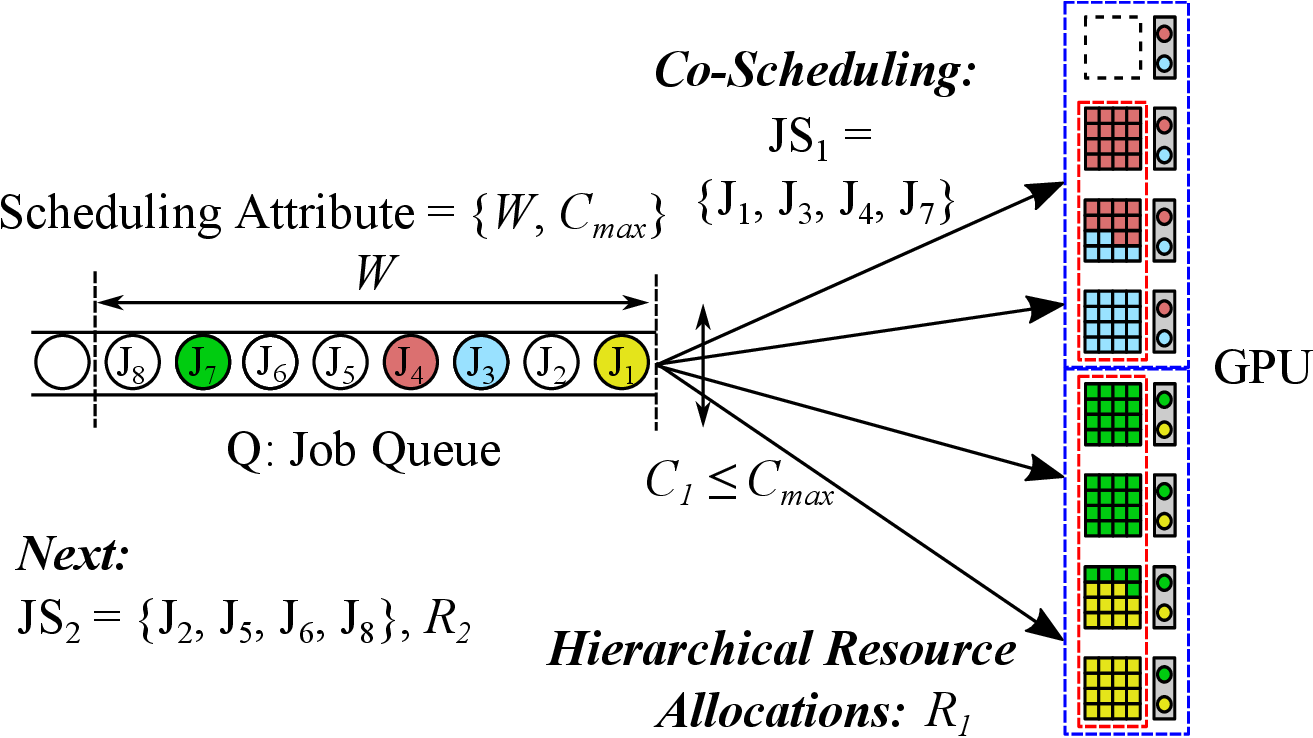}
  \caption{Co-Scheduling and Resource Partitioning Problem}
  \label{fig:problem}
\end{center}
\end{figure}

\subsection{Problem Definition}\label{prob-def}
Figure~\ref{fig:problem} illustrates the optimization problem we solve in this paper. 
We target the first $W$ jobs in the job queue ($\mathrm{Q = \{J_1, J_2, \cdots, J_W\}}$) for the co-scheduling and resource partitioning decision making. 
We then choose a set of jobs to co-schedule ($\mathrm{JS_1 = \{J_1, J_3, J_4, J_7\}}$) and decide the resource partitioning and allocations (denoted as $R_1$). 
Here, the number of co-located jobs or concurrency ($C_1$) is constrained by a parameter $C_{max}$. 
After this optimization procedure, we eventually obtain sets of decisions: (1) a set of co-scheduling sets denoted as $\mathrm{L_{JS}} = \{\mathrm{JS_1}, \mathrm{JS_2}, \cdots \}$; and (2) a set of corresponding resource allocations denoted as $\mathrm{L_{R}} = \{R_1, R_2, \cdots \}$. 

This optimization procedure is formulated as follows: 
\begin{eqnarray}
    &given & \quad W, \quad C_{max}, \quad \mathrm{Q = \{J_1, J_2, ... , J_W\}} \nonumber\\
    &min & \quad \sum_{i=1}^{|\mathrm{L_{JS}}|} CoRunTime(\mathrm{JS_{i}}, R_{i}) \nonumber\\
    &s.t. & \quad CoRunTime(\mathrm{JS_{i}}, R_{i}) \le SoloRunTime(\mathrm{JS_{i}}) \nonumber\\
    &&\quad 1 \le C_i (=|\mathrm{JS_{i}}|) \le C_{max} \nonumber\\
    &&\quad \forall i \in [1, |\mathrm{L_{JS}}|], \hspace{20pt} |\mathrm{L_{JS}}| = |\mathrm{L_{R}}| \nonumber\\
    &&\quad \mathrm{JS_1 \cup ... \cup JS_{|L_{JS}|} = Q} \nonumber\\
    &&\quad \mathrm{|JS_1| + ... + |JS_{|L_{JS}|}|} = W \nonumber\\
    &output & \quad \mathrm{L_{JS} = \{JS_1, JS_2, ...\}}, \quad \mathrm{L_{R}} = \{R_1, R_2, ...\} \nonumber
\end{eqnarray}



\begin{table}[b]
\caption{Definitions of Parameters/Functions}
\label{table:refs}
{
\scriptsize
\centering
\begin{tabular}{|M{0.28\linewidth}||M{0.62\linewidth}|}
    \hline
    Parameter or Function & Remarks \\
    \hline\hline
    $\mathrm{Q}$ & Queuing jobs within the window: $\mathrm{Q} = \{\mathrm{J_1}, \mathrm{J_2}, \cdots, \mathrm{J_{W}}\}$ \\
    \hline
    $\mathrm{J_i}$ & $i$th job in the queuing jobs \\
    \hline
    $W$ & The number of jobs within the window on the queue  \\
    \hline
    $C_{max}$ & The maximum number of concurrently executed jobs \\
    \hline
    $\mathrm{L_{JS}}$ & A list of job sets to be co-scheduled: $\mathrm{L_{JS}} = \{\mathrm{JS_1}, \mathrm{JS_2}, \cdots\}$ \\
    \hline
    $\mathrm{JS_i}$ & $i$th set of jobs in $\mathrm{L_{JS}}$ to be co-scheduled \\
    \hline
    $\mathrm{L_{R}}$ & A list of resource partitioning/allocation setups associated with the job sets: $\mathrm{L_{R}} = \{R_1, R_2, \cdots\}$ \\
    \hline      
    $R_i$ & The resource partitioning/allocations for $\mathrm{JS_i}$ \\
    \hline
    $C_{i}$ ($= |\mathrm{JS_{i}}|$) & The concurrency of $i$th co-scheduled job set \\
    \hline
    $CoRunTime(\mathrm{JS_i}, R_i)$ & The total execution time when co-locating $\mathrm{JS_i}$ w/ $R_i$ \\
    \hline    
    $SoloRunTime(\mathrm{JS_i})$ & The total time when executing $\mathrm{JS_i}$ w/ time sharing\\
    \hline

\end{tabular}
}
\end{table}

We solve this throughput-oriented optimization problem where we minimize the overall co-run execution time (\textit{CoRunTime}) for the sets of selected jobs ($\mathrm{L_{JS}}$) and corresponding hardware configurations ($\mathrm{L_{R}}$). 
The scheduling attributes ($\{W, C_{max}\}$) and the queuing jobs ($\mathrm{Q}$) are given.
The goal is to find the optimal set of co-scheduling job-sets ($\mathrm{L_{JS}}$) as well as their associated resource allocations ($\mathrm{L_{R}}$), and thus they are the outputs. 
In this context, \textit{optimal set of co-scheduling job-sets} refers to the selection of compatible job-sets from the given job window, which maximize the overall co-run throughput. 
The first constraint represents that co-scheduling $i$th set of jobs in $\mathrm{L_{JS}}$ must improve performance compared with the time-shared scheduling, i.e., running the jobs one by one with using the entire GPU resources exclusively. 
The second constraint restricts the co-scheduling concurrency, i.e., the concurrency ($C_i$) must be less than or equal to the given upper limit ($C_{max}$). These two constraints stand for any $i$ ($1\leq i \leq \mathrm{|L_{JS}|}$). 
The last two constraints restrict the job set selections, i.e., they are selected from the queue ($\mathrm{Q}$) in a mutually exclusive and collectively exhaustive manner. 
The parameters and functions used in this optimization procedure are listed in Table~\ref{table:refs}.

\begin{figure}
\includegraphics[width=\linewidth]{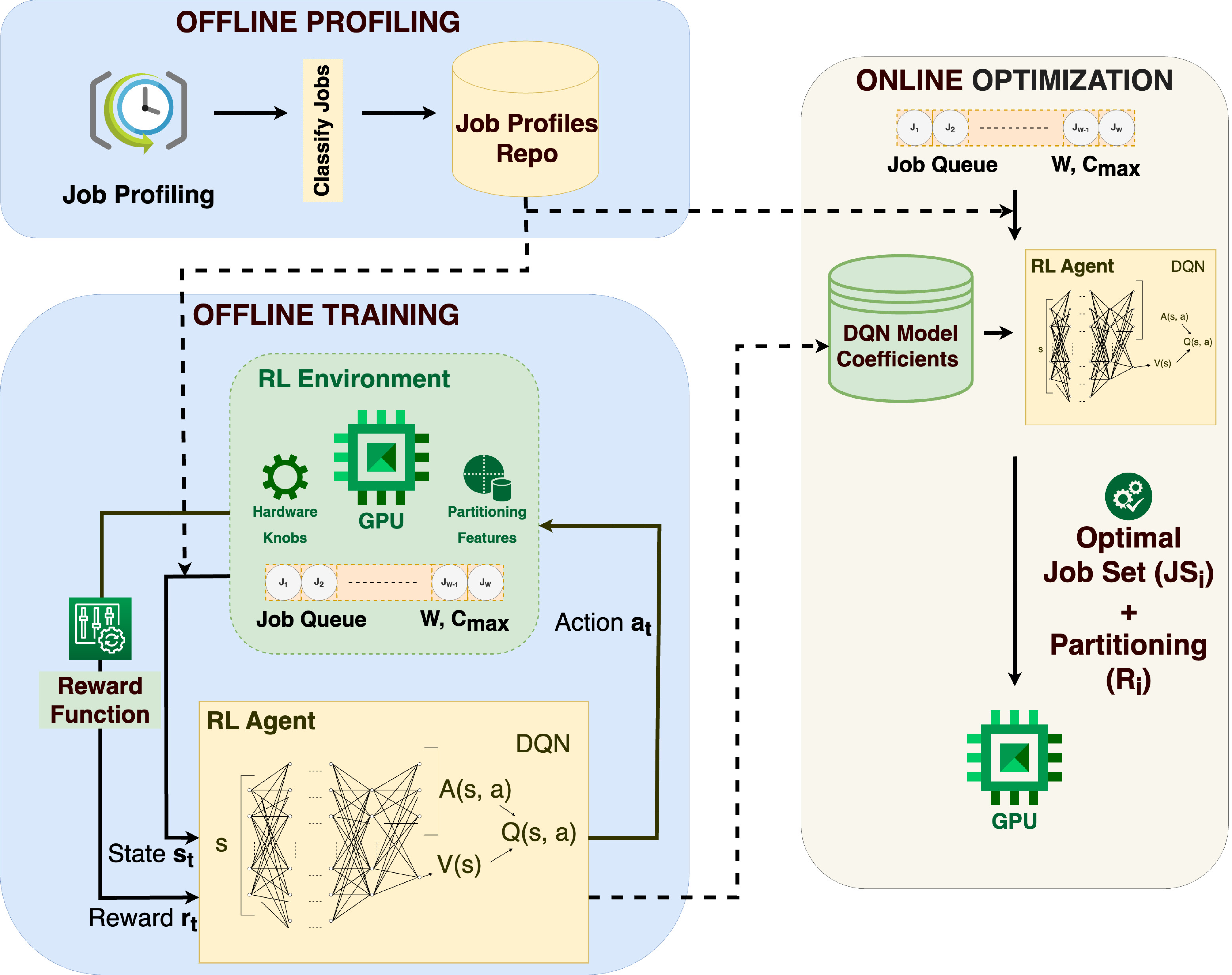}
  \caption{Proposed System Architecture}
  \label{fig:solution-overview}
\end{figure}

\subsection{System Design}


\figurename{ \ref{fig:solution-overview}} illustrates the entire system architecture of our solution. 
As shown in the figure, the overall solution consists of three parts: (1) the offline profiling to collect application profiles; (2) the offline training to setup the coefficients of our agent; and (3) the online optimization to apply the trained agent to the decision making. 

For the application profiling, we collect hardware performance counters to characterize the running jobs on the target system. The exact counter selections are listed on Table~\ref{hw-counters} in Section~\ref{setup}. 
The profiles need to be collected beforehand for any co-scheduling targets in both the offline and online phases. 
In the offline training phase, we collect the solo-run profiles for all the benchmark programs before the model training. 
In the online optimization phase, if no profile is available for a queuing job, it is excluded from the co-scheduling target and is executed with exclusively using the entire GPU while collecting the profile that shall be stored in the \textit{Job Profiles Repository}. 
If the application is executed again on the system, it is included in the co-scheduling target as the profile is available in the repository. 
This procedure requires a matching function to select a corresponding profile for each job based on its submission information (e.g., binary path, user ID, etc.). 
In this study, we simply consider using the application binary path plus name as a key and checking if there is a profile associated with it in the repository. 
Developing an advanced way to generate the key from the job submission information, while taking a variety of aspects into account (e.g., input dependency\footnote{For instance, the characteristics/behavior of an application can depend on its inputs, and there are several promising solutions to compensate for it~\cite{input-dependency}. }), is an open problem for profile-based approaches in general, and our matching function will be replaced with a more sophisticated scheme in our future work. 
 


For the offline model training, we create variants of benchmark program mixes to co-locate on the target GPU. 
For each program mix, we continuously examine the co-run throughput while changing the partitioning setup. 
This partitioning search is based on reinforcement learning, i.e., we update the partitioning and resource allocations accordingly when the next co-run (with the exact same program mix) based on the reward function output that takes the co-run throughput into account. 
During this procedure, the state-action table, which is approximated by a neural network in this study, is trained, and the model coefficients in the agent are eventually determined. 
The model coefficients are hardware specific and are not portable to different hardware, however the training procedure is required only once for a system though.

We take this offline training approach based on reinforcement learning due to the following reasons. 
First, the resource partitioning setup is \textit{not} configurable dynamically at runtime in commercial GPUs, and thus we cannot adaptively/dynamically learn the optimal configurations for a given set of jobs in the queue ($\mathrm{Q}$) by testing various configurations at runtime. 
Second, in the offline training phase, we apply reinforcement learning instead of utilizing well-known supervised learning using training dataset because it is infeasible to obtain the \textit{labeled} dataset. 
More specifically, labeling here associates a given job mix with the \textit{best} co-scheduling and resource partitioning decisions, which requires the \textit{exhaustive search} for each job mix (or data) in the dataset.


In the online phase, we deploy an optimization agent that solves the optimization problem formulated in Section~\ref{prob-def} using the model generated in the offline phase. 
The agent regards the optimization as a classification problem and uses the model to choose sets of co-scheduling job mixes ($\mathrm{L_{JS}}$) and corresponding resource allocations ($\mathrm{L_{R}}$) to maximize the GPU throughput. 
In this work, we do not update the model during the online phase, however dynamically refining the trained model is a promising option for our future work.

\subsection{Reinforcement Learning-based Solution}
In reinforcement learning, an agent learns what action to take based on the situation so as to maximize the cumulative reward \cite{b8}. 
The goal of this form of learning is to enable an agent to explore the parameter space based on its interaction with the environment, perform trial-and-error, and eventually generalize to perform optimal set of actions to reach the goal state. 
The properties of reinforcement learning relevant to this work are as follows:

\subsubsection{Agent}
An agent is an entity that interacts with the environment, receives feedback (reward signal), and learns a policy that governs the behavior at a given state of the system. It learns an optimal policy in order to maximize the accumulation of the reward signals in the offline training in our approach. In this work, our agent serves as a co-scheduler that selects the sets of job mix and the associated partitioning ($\mathrm{L_{JS}}$ and $\mathrm{L_{R}}$) from the given queue ($\mathrm{Q}$). 

\subsubsection{Environment}
An environment acts as a black-box for the agent. 
In this work, the environment consists of the target GPU and its hardware features. 


\subsubsection{State}
The representation of the current situation of the system is defined as the state. The state should contain all the relevant information required for deciding the actions. In our approach, the state of the system represents all the jobs in the current job window ($\mathrm{Q} = \{\mathrm{J_1}, \mathrm{J_2}, \cdots, \mathrm{J_{W}}\}$) along with their job features characterized by their profiles.

\subsubsection{Action}
An action is a decision made by the agent based on the current state of the system. 
For our approach, actions can include decisions for selecting the sets of co-scheduling job mix and corresponding resource allocation ($\mathrm{L_{JS}}$ and $\mathrm{L_{R}}$). 

\subsubsection{Reward}
A reward signal define the goal of the reinforcement learning~\cite{b8}. 
For every action, the agent receives the reward signal as a numerical value. 
As agent's goal is to maximize the cumulative reward, the reward signal quantifies and evaluates an action at a given state of the system. 
The details of the setup for this reward function will be provided in Section~\ref{setup}. 


\subsection{Agent Implementation with Deep Q-Learning}\label{dql}

We apply deep Q-learning to the offline training for optimizing the actions, i.e., co-scheduling and resource partitioning decisions, made by the agent.  
For a given finite Markov Decision Process, Q-learning can be used to determine the optimal \textit{Q-value function}. For a given state $s$ and action $a$, $Q(s, a)$ (Q-value function) can be defined as the \textit{expected value} of the overall rewards. The optimal Q-value function (also known as action-value function) has been defined using Bellman Optimality Equation~\cite{b9}.
\begin{eqnarray}
Q^*(s, a) = \mathbb{E}[I_s^a + \gamma\sum_{s' \in S} max Q^*(s', a')]\nonumber
\end{eqnarray}
In this formulation, there is an immediate reward $I_s^a$ which will be the gain for taking the action $a$ at the state $s$ and there is a long-term value which is an estimate of the values of the series of actions and state transitions. $\gamma$ is the discount factor which defines the weight for the long-term rewards. The Q-values for the given state-action pair are updated as per the following update rule: $Q^{new}(s_t, a_t) \leftarrow Q(s_t, a_t) + \alpha (r_t + \gamma \max_a Q(s_{t+1}, a) - Q(s_t, a_t))$ where $\alpha$ and $\gamma$ are the learning rate and the discount factor respectively. Conventionally, the Q-value function has been estimated by generating the Q-table, for mapping every state-action pairs.

For more complex and higher dimensional state spaces, it might not be possible to estimate the optimal values using the Q-table and hence deep neural networks would be useful. As neural networks are non-linear function approximators, they are well-suited to estimate the optimal action-value function in the process of Q-learning. In particular, in this work, we use 
a \textit{duelling double deep Q network}. The choice of this network is based on the benefits highlighted from two separate works by Hasselt et al.~\cite{b10} and Wang et al.~\cite{b11}.

\begin{table}[b]
{
\scriptsize
\caption{Evaluation Environment}\label{environment}
\vspace{-10pt}
\begin{center}
\begin{tabular}{ |M{0.18\linewidth}||M{0.7\linewidth}| } 
\hline
 Name & Remarks \\\hline\hline
 GPU &  NVIDIA A100 40GB PCIe 250W TDP \\\hline
 Operating System & Ubuntu 20.04.4 LTS, Kernel Version: 5.4.0-137-generic \\\hline
 Software & CUDA Version: 11.6, Driver Version: 510.108.03, Python Version: 2.7.18 \\\hline
\end{tabular}
\end{center}
}
\end{table}

\section{Evaluation}\label{Evaluation}
In Section~\ref{setup}, we first describe our evaluation setups including our platform, workload selections, neural network configurations, compared methods, and partitioning variants. 
We then introduce our evaluation results in Section~\ref{result}.
\subsection{Evaluation Setup}\label{setup}

\begin{table}[b]
{
\scriptsize
\caption{Collected Hardware Performance Counters}\label{hw-counters}
\vspace{-10pt}
\begin{center}
\begin{tabular}{|M{0.94\linewidth}|}
\hline
\textbf{Statistics} \\\hline\hline
Duration, Memory [\%], Elapsed Cycles, Grid Size, Registers Per Thread, DRAM Throughput, L1/TEX Cache Throughput, L2 Cache Throughput, SM Active Cycles, Compute (SM) [\%], Waves Per SM,  Achieved Active Warps Per SM \\\hline
\end{tabular} 
\end{center}
}
\end{table}

\subsubsection{Platform}\label{setup-platform}
Table~\ref{environment} lists the system environment used for evaluating our approach. As mentioned before, we utilized an A100 GPU and applied the MIG and MPS features to it. 
Our system is implemented in Python using multiple standard python libraries. We build our reinforcement learning environment using the gymnasium python library~\cite{gym-doc}. For implementing the agent, we use the PyTorch library for implementing the deep neural networks for Q-learning~\cite{pytorch}. Further, we use scikit-learn for performing additional data pre-processing and feature engineering~\cite{scikit-learn}.
We collect hardware performance counters to profile and characterize the applications. 
To this end, we utilize the NVIDIA Nsight compute framework~\cite{nsight}. 
Table~\ref{hw-counters} lists the collected hardware performance counters by using the framework. 
These statistics are useful to characterize the applications in terms of compute intensity, memory intensity, parallelism/scalability, memory access pattern, and so forth. 

\begin{table}[b]
{
\scriptsize
\caption{Benchmark Classifications}\label{wl-class}
\vspace{-10pt}
\begin{center}
\begin{tabular}{|M{0.18\linewidth}||M{0.7\linewidth}|}
\hline
Class & Benchmarks \\\hline\hline
CI & lavaMD, huffman*, hotspot3D, hotspot*, heartwall*, bt\_solver\_A, bt\_solver\_B, bt\_solver\_C\\\hline
MI & lud\_A, lud\_B, lud\_C*, sp\_solver\_A, sp\_solver\_B, sp\_solver\_C, randomaccess, cfd*, gaussian*, stream\\\hline
US & kmeans, dwt2d, needle*, pathfinder, backprop*, qs\_Coral\_P1, qs\_Coral\_P2, qs\_NoFission*, qs\_NoCollisions\\\hline
\end{tabular} 

\end{center}
}
\end{table}

\subsubsection{Workloads}\label{setup-workloads}
We utilize the \texttt{Rodinia} benchmark suite~\cite{rodinia}, a \texttt{stream} benchmark~\cite{stream}, a \texttt{randomaccess} benchmark~\cite{random}, and the \texttt{Quicksilver} mini application chosen from the CORAL benchmark suite~\cite{coral}. 
These benchmark programs are classified into \textit{CI (Compute Intensive)}, \textit{MI (Memory Intensive)}, and \textit{US (UnScalable)} as shown in Table~\ref{wl-class}. 
We follow a prior study for the classification procedure~\cite{gpu-mig-power}: (1) if the performance degradation caused by 1GPC run with the private memory option compared with the full 8GPC run is less than 10\%, we regard it as an UnScalable (US) application; (2) otherwise, if the ratio of \texttt{Compute (SM) [\%]} to \texttt{Memory [\%]} is more than 0.80, we regard it as a CI application; (3) otherwise it is an MI application.

\begin{table}[!b]
\begin{center}
\centering
\caption{Tested Job Mixes per Category ($W=12$)}
\label{job-mix}
\scriptsize
\begin{tabular}{|M{0.15\linewidth}|M{0.07\linewidth}||M{0.62\linewidth}|}
\hline
Category & Name & Jobs \\\hline\hline
& Q1 & huffman*, bt\_solver\_C, bt\_solver\_B, hotspot3D, heartwall*, lavaMD, lud\_B, cfd*, sp\_solver\_B, pathfinder, needle*, qs\_NoFission*\\\cline{2-3} 
CI-dominant & Q2 & bt\_solver\_C, heartwall*, lavaMD, huffman*, hotspot*, hotspot3D, cfd*, sp\_solver\_C, gaussian*, pathfinder, needle*, qs\_Coral\_P1\\\cline{2-3} 
(CIx6, MIx3, USx3) & Q3 & huffman*, bt\_solver\_C, hotspot3D, hotspot*, heartwall*, lavaMD, lud\_B, stream, sp\_solver\_C, qs\_NoFission*, pathfinder, needle*\\\hline
& Q4 & bt\_solver\_B, heartwall*, bt\_solver\_C, lud\_B, gaussian*, sp\_solver\_B, cfd*, sp\_solver\_C, stream, qs\_NoCollisions, pathfinder, qs\_Coral\_P2 \\\cline{2-3}
MI-dominant & Q5 & heartwall*, hotspot*, bt\_solver\_B, lud\_B, gaussian*, randomaccess, stream, lud\_C*, sp\_solver\_B, qs\_Coral\_P2, dwt2d, qs\_Coral\_P1 \\\cline{2-3} 
(CIx3, MIx6, USx3) & Q6 & bt\_solver\_C, huffman*, lavaMD, sp\_solver\_B, gaussian*, randomaccess, lud\_C*, stream, cfd*, qs\_NoFission*, needle*, qs\_Coral\_P1 \\\hline
& Q7 & heartwall*, hotspot*, hotspot3D, gaussian*, stream, lud\_B, pathfinder, qs\_NoFission*, qs\_Coral\_P2, backprop*, qs\_NoCollisions, dwt2d \\\cline{2-3}
US-dominant & Q8 & bt\_solver\_C, hotspot3D, lavaMD, stream, cfd*, lud\_B, qs\_Coral\_P1, needle*, kmeans, qs\_Coral\_P2, qs\_NoFission*, qs\_NoCollisions \\\cline{2-3} 
(CIx3, MIx3, USx6) & Q9 & lavaMD, hotspot3D, hotspot*, sp\_solver\_B, lud\_C*, randomaccess, qs\_Coral\_P1, dwt2d, kmeans, needle*, qs\_NoCollisions, qs\_Coral\_P2 \\\hline
& Q10 & lavaMD, huffman*, hotspot3D, bt\_solver\_C, lud\_C*, lud\_B, stream, sp\_solver\_C, qs\_NoCollisions, needle*, pathfinder, qs\_Coral\_P1 \\\cline{2-3}
Balanced & Q11 & huffman*, hotspot3D, hotspot*, bt\_solver\_B, cfd*, lud\_C*, stream, gaussian*, qs\_Coral\_P2, needle*, pathfinder, dwt2d \\\cline{2-3} 
(CIx4, MIx4, USx4) & Q12 & lavaMD, hotspot*, huffman*, heartwall*, sp\_solver\_C, lud\_C*, randomaccess, gaussian*, needle*, pathfinder, qs\_NoCollisions, backprop* \\\hline
\end{tabular}

\end{center}
\end{table}

In our evaluation, we first setup the job window size ($W$) to twelve. We later scale the size as well to assess the impact of the window size selection. 
For the offline training, we exclude nine programs marked with * in the Table~\ref{wl-class} and use the remaining 18 programs. 
The objective of the exclusion procedure is to check if our approach can generalize to unseen applications.  
We create 20 different job queues for the agent training, each of which consists of $W$ programs randomly selected from the 18 programs while including all the 3 categories in the queue. 
As for the online inference, we test our approach with using different types of job mixes: (1) \textit{CI-dominant}; (2) \textit{MI-dominant}; (3) \textit{US-dominant}; and (4) \textit{Balanced}. 
On one hand, the \textit{X-dominant} job mix is composed of 50\% of X class applications (X=CI, MI, or US), and the rest of the 50\% are from the other classes selected in a round robin manner. 
For instance, when $W=12$, the \textit{CI-dominant} class consists of 6CI, 3MI, and 3US applications. 
On the other hand, the \textit{Balanced} job mix selects a set of application classes in a round robin manner, and when $W=12$, it consists of 4CI, 4MI, and 4US applications. 
For each of these job mix categories, we create several job mix variants (A, B, and C), and for each job mix variant, we assign applications to each application class, which are randomly selected by using Table~\ref{wl-class}. 
The exact job mix selections for $W=12$ are listed in Table~\ref{job-mix}. 
Note the programs marked with * are unseen in the training.

\subsubsection{Setups for Training and Inference}\label{training-setup}


Table~\ref{nn-params} lists the setups used for the reward function and the agent.  
In this evaluation, we use two kinds of rewards: (i) intermediate reward $r_i$ and (ii) final reward $r_f$. 
On one hand, the intermediate reward evaluates the resource allocation for a selected job, which can be assessed before launching the job using the associated profile. 
It returns a higher reward when assigning a resource where it is needed (e.g., allocating more memory bandwidth to an memory intensive application). 
On the other hand, the final reward refers to the measured throughput improvement over the time-sharing executions, which is obtained only after the completion of co-running a job mix. 

In the table, $SmAllocRatio$ and $MemoryAllocRatio$ are hardware parameters, which characterize (i) the ratio of allocated Streaming-Multiprocessors to the total number of them, and (ii) the ratio of allocated memory bandwidth to the total available memory bandwidth respectively. 
$ComputeRatio$, $MemoryRatio$ and $DurationRatio$ are job-specific profile parameters which are described as follows: (i) $ComputeRatio$: the ratio of \texttt{Compute (SM) [\%]} of the current job to the mean \texttt{Compute (SM) [\%]} of the job window, (ii) $MemoryRatio$: the ratio of \texttt{Memory [\%]} of the current job to the mean \texttt{Memory [\%]} of the job window, and (iii) $DurationRatio$: the ratio of solo-run execution time of the current job to the mean solo-run execution time of the job window. 
With this particular formulation of the reward function, our focus has been on optimizing for co-run throughput. Nevertheless, this approach can be further expanded by fine-tuning the reward function to encompass additional parameters, including job-specific priorities, scheduling fairness and energy consumption.

As for the agent, it is configured with double dueling deep Q-network~\cite{b11}, and the details are listed also in Table~\ref{nn-params}. 
In a dueling deep Q-network, the Q-value is split into two values: (i) $V$ value of being in the given state, and (ii) $A$ advantage of selecting a particular action in the given state. 
More details about the update rule for Q-value, with use of A and V can be seen in the work by Wang et al.~\cite{b11}. 
Further, by following the existing work~\cite{b10}, we use two different networks based on the same described architecture: one for predicted Q-value and the other for target Q-value. 
For the training, we use the well-known $\epsilon$-greedy approach, in which we initially set a parameter $\epsilon$ to 1 and gradually decrease it until it reaches a certain point (e.g., 0.01 in our evaluation). 
The parameter $\epsilon$ controls the frequency of random actions taken by the agent. 
More specifically, with a probability of $\epsilon$, the agent takes an action randomly chosen from the entire search space. 
This procedure is meant to converge to the global optimal as far as possible. 
After the training procedure is completed, we set the $\epsilon$ to 0 so as not to take any random action when using the trained agent in the online phase.

\begin{table}[b]
{
\scriptsize
\caption{Agent and Reward Function Setups}\label{nn-params}
\vspace{-10pt}
\begin{center}
\begin{tabular}{ |M{0.12\linewidth}||M{0.76\linewidth}| } 
\hline
Type & Setups
\\\hline\hline
Reward Function & $r_i = (SmAllocRatio \times ComputeRatio + MemoryAllocRatio \times MemoryRatio) \times DurationRatio^2$
$r_f = (SoloRunTime/CoRunTime - 1) \times 100 $
\\\hline
Agent & \lbrack\textbf{\# of neurons in the input layer}\rbrack: $W \times (f + 5)$, 
\lbrack\textbf{\# of neurons in the output layer}\rbrack: $V$ = 1, $A$ =29,
\lbrack\textbf{\# of hidden layers}\rbrack: 3, \lbrack\textbf{\# of neurons in each hidden layer}\rbrack: 512/256/128, 
\lbrack\textbf{Layer NW}\rbrack: Fully connected, \lbrack\textbf{Activation function}\rbrack: Rectified Linear
\\\hline 
\end{tabular}
\end{center}
}
\end{table}

\subsubsection{Compared Methods}
To assess the effectiveness of our approach, we compare the following different scheduling policies. We compare them in terms of  throughput, application performance, and fairness when scheduling given job mixes. 
\begin{itemize}
\item \textbf{Time Sharing (Baseline)}: Jobs in the given job mix (or queue) are executed using the entire GPU resources exclusively without co-scheduling/partitioning. 
\item \textbf{MIG Only ($C=2$)}: Following the existing studies~\cite{gpu-mig-power,cos-hetero2}, we test a MIG only option with the concurrency $C$ at 2. The job set selections and assignments are optimal, i.e., exhaustively chosen from all the possible setups.
\item \textbf{MPS Only ($C\leq C_{max}$)}: We test the MPS only option with concurrency selections ($C\leq C_{max}$). The job set selections and resource assignments are determined through an exhaustive search too. 
\item \textbf{MIG+MPS Default ($C\leq C_{max}$)}: The MIG partitioning is selected so that the average throughput across Q1-Q12 is maximized. The MPS allocation is set to the \textit{default} mode.
The job set selections ($L_{JS}$) are optimal, i.e., they are chosen through an exhaustive search within the designated concurrency limit and configuration space. 
\item \textbf{MIG+MPS w/ RL ($C\leq C_{max}$)}: Our proposed reinforcement learning-based co-optimization of co-scheduling and hierarchical partitioning. 
\end{itemize}

\begin{table}[b]
{
\scriptsize
\caption{Partitioning Setups for Different Concurrency (See Section~\ref{evaluated-partitions} for the Format Definition)}\label{part-select}
\begin{center}
\begin{tabular}{ |M{0.02\linewidth}||M{0.35\linewidth}|M{0.47\linewidth}| } 
\hline
$C$ & For MPS Only & For MPS+MIG w/ RL \\\hline\hline
 2 & [(0.1)+(0.9),1m]; [(0.2)+(0.8),1m]; \dots; [(0.5)+(0.5),1m]; & [(0.1)+(0.9),1m]; [(0.2)+(0.8),1m]; \dots; [(0.5)+(0.5),1m]; [\{0.375\}+\{0.5\},1m]; [\{0.375\},0.5m]+[\{0.5\},0.5m] \\\hline
 3 & [(0.1)+(0.1)+(0.8),1m]; \dots; [(0.34)+(0.33)+(0.33),1m]; & [(0.1)+(0.1)+(0.8),1m]; \dots; [(0.34)+(0.33)+(0.33),1m]; [\{0.375\},0.5m]+[(0.1)+(0.9),\{0.5\},0.5m]; \dots; [\{0.375\},0.5m]+[(0.5)+(0.5),\{0.5\},0.5m]; [\{0.375\}+(0.1),(0.9)\{0.5\},1m]; \dots; [\{0.375\}+(0.5),(0.5)\{0.5\},1m]; \\\hline
 4 & [(0.1)+(0.1)+(0.1)+(0.7),1m]; \dots; [(0.25)+(0.25)+(0.25)+(0.25),1m]; & [(0.1)+(0.1)+(0.1)+(0.7),1m]; \dots; [(0.25)+(0.25)+(0.25)+(0.25),1m]; [(0.1)+(0.9),\{0.375\},0.5m]+ [(0.1)+(0.9),\{0.5\},0.5m]; \dots; [(0.5)+(0.5),\{0.375\},0.5m]+ [(0.5)+(0.5),\{0.5\},0.5m]; [(0.1)+(0.9)\{0.375\}+(0.1)+(0.9)\{0.5\},1m]; \dots; [(0.5)+(0.5)\{0.375\}+(0.5)+(0.5)\{0.5\},1m]; \\\hline

\end{tabular}
\end{center}
}
\end{table}

\subsubsection{Evaluated Partitions}\label{evaluated-partitions}
Table~\ref{part-select} lists all the partitioning variants explored in the evaluation for different concurrency setups ($C$). We list them for \textit{MPS Only} and \textit{MIG+MPS w/ RL} described above. 
For \textit{MIG Only}, we explore the two options shown in Figure~\ref{partitioning-variants} to compare with the existing works~\cite{gpu-mig-power, cos-hetero2}. 
For \textit{MIG+MPS w/ Default}, it assigns the default active thread percentage over the optimized MIG partitions.

The format to represent partitioning states is defined as follows. 
First, a GI or the entire GPU is enclosed in a square brackets. It is denoted as [compute resource setup, assigned memory resource]. For the memory resource part, when $\alpha\times100$\% of the entire GPU memory bandwidth is assigned, it is denoted as "$\alpha$m". 
As for the compute resource setup, a CI or an MPS process is enclosed in curly brackets or parentheses, respectively. 
The number in brackets (let it be $\beta$) represents the amount of allocated compute resources (i.e., $\beta\times100$\% of the GPU total). For instance, [\{$\beta$\}, $\alpha$m] shows one CI exists inside the GI, which can utilize $\beta\times100$\% (or $\alpha\times100$\%) of compute (or bandwidth) resources. 
Further, the partitions in the same level of the hierarchy are combined with "+" in the format. 
For instance, [\{0.375\}+\{0.5\},1m] is the 3GPC+4GPC MIG-only partitioning with the shared memory option, whereas [\{0.375\},0.5m]+[\{0.5\},0.5m] is the private memory option with the same GPC allocations.

\subsection{Experimental Results}\label{result}

\begin{figure}[t]
  \centering
  \includegraphics[height=5.5cm]{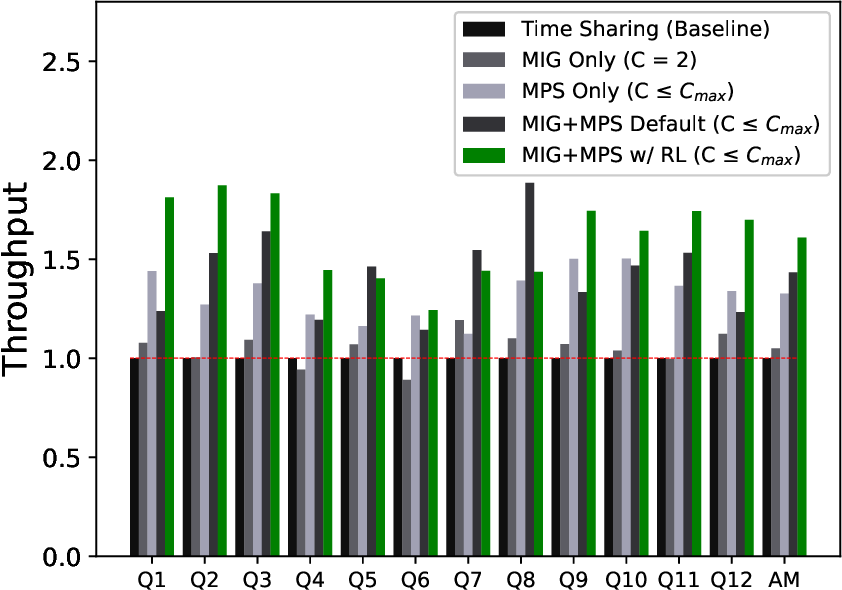}
  \caption{Throughput Comparison ($C_{max}=4$, $W=12$)}
  \label{fig:tpt}
\end{figure}

Figure~\ref{fig:tpt} compares throughput among different methods and across different workloads. The horizontal axis represents executed workloads (AM: Arithmetic Mean), while the vertical axis indicates relative throughput normalized to that of \textit{Time Sharing} for each workload. 
Throughout the evaluation, the maximum concurrency ($C_{max}$) is set at $4$. 
In general, the proposed reinforcement learning-based approach outperforms all the other methods for almost all the workloads. 
Compared with the \textit{Time Sharing}, it achieves 1.516 or 1.873 times throughput improvement on average or at best, respectively. 
The \textit{MIG+MPS Default} is also hierarchical with a constant MIG partitioning and the default MPS setup. 
Our approach outperforms this option, which implies that the hierarchical partitioning needs to be changed depending on the characteristics of jobs to be co-located. 
The \textit{MPS Only} option is less effective than ours because it is not capable of mitigating the interference on the shared resources among co-scheduled programs. 
By combining with the MIG feature, it becomes even more effective. 

\begin{figure}[t]
\centering
  \includegraphics[height=4cm]{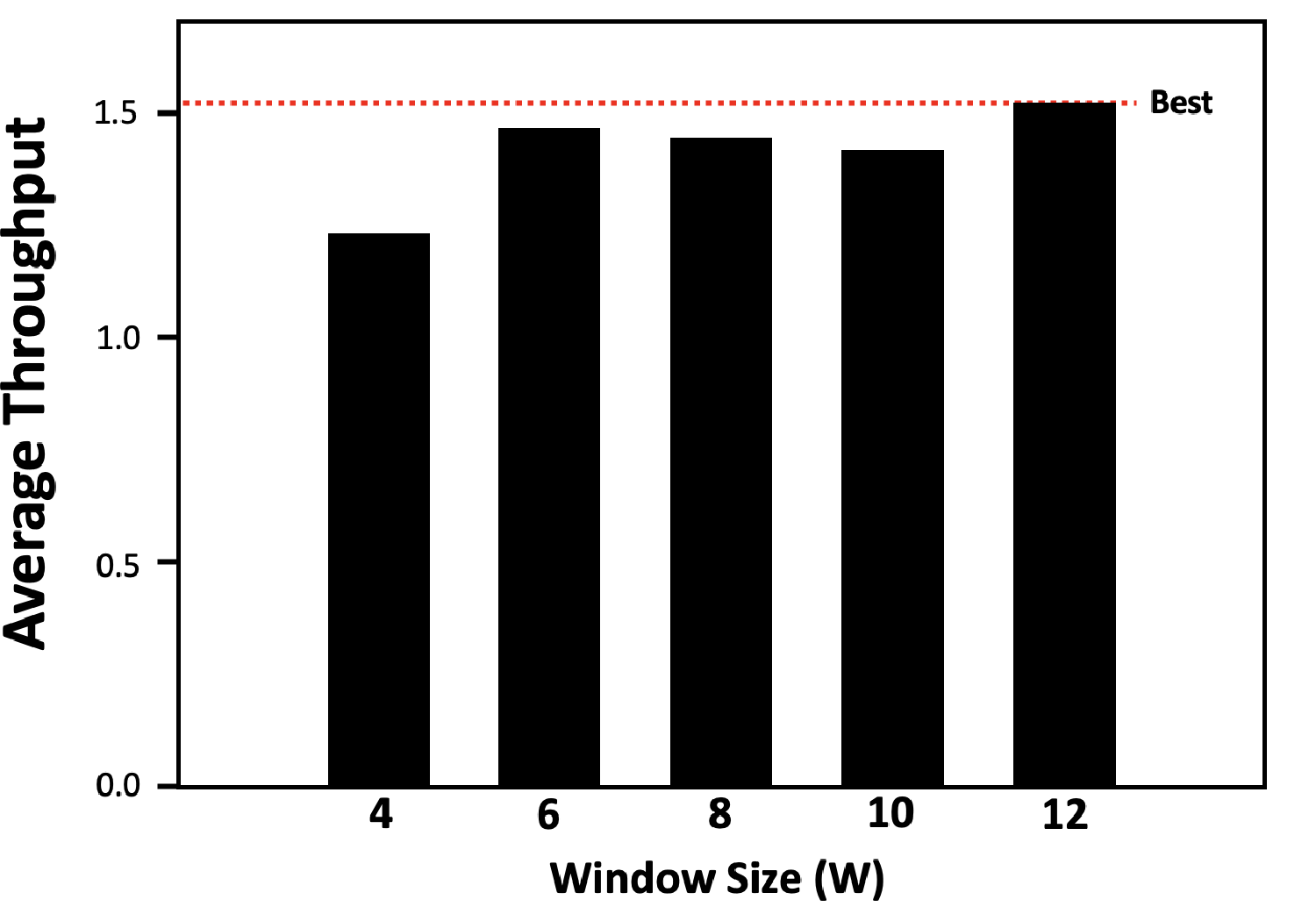}
  \caption{Average Throughput Comparison for various Window Sizes ($C_{max}=4$)}
  \label{fig:w_scale}
\centering
  \includegraphics[height=4cm]{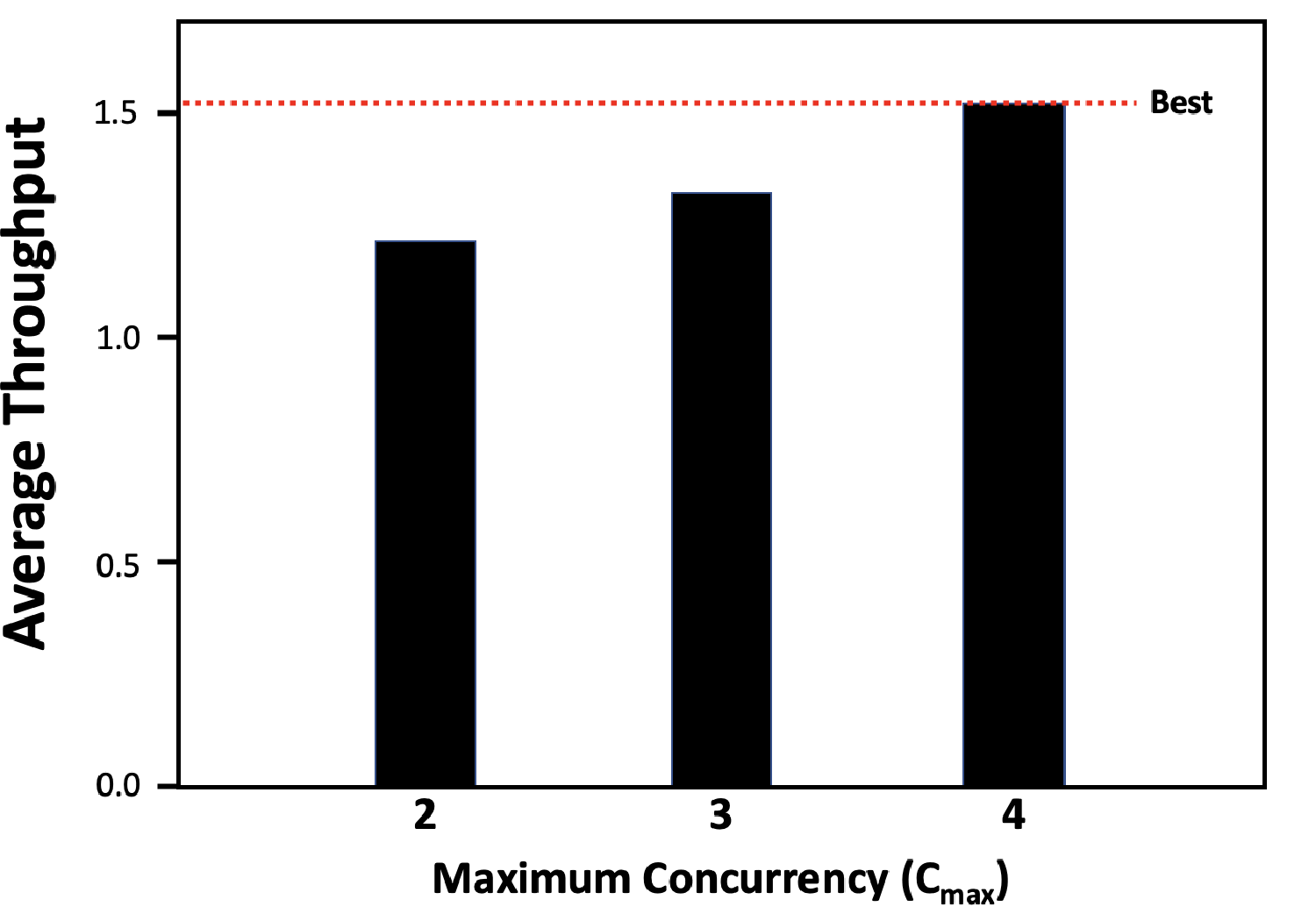}
  \caption{Average Throughput Comparison for various values of $C_{max}$ ($W=12$)}
  \label{fig:c_scale}
\end{figure}

Next, Figure~\ref{fig:w_scale}~and Figure~\ref{fig:c_scale} present the average throughput as a function of the window size ($W$) and the maximum job concurrency ($C_{max}$). 
The vertical axes of the mentioned figures represent the average throughput based on all of the 12 job queues, and the horizontal axes represent $W$ and $C_{max}$ respectively. 
Note that $C_{max}$ is fixed at 4 when scaling $W$ in Figure~\ref{fig:w_scale}, while $W$=12 stands when scaling $C_{max}$ in Figure~\ref{fig:c_scale}. 
As shown in the figures, the throughput increases as we scale these parameters. 
This is because of the following reasons: (1) our approach can find better co-scheduling groups for higher $W$; and (2) our co-scheduling can utilize resources more effectively for higher $C_{max}$ thanks to the flexible partitioning and shared resource isolation features offered by MPS and MIG. 
We selected $W$=12 and $C_{max}$=4 as scaling them further did not improve the throughput further for our workloads. 


\begin{figure}[t]
\centering
  \includegraphics[height=5.5cm]{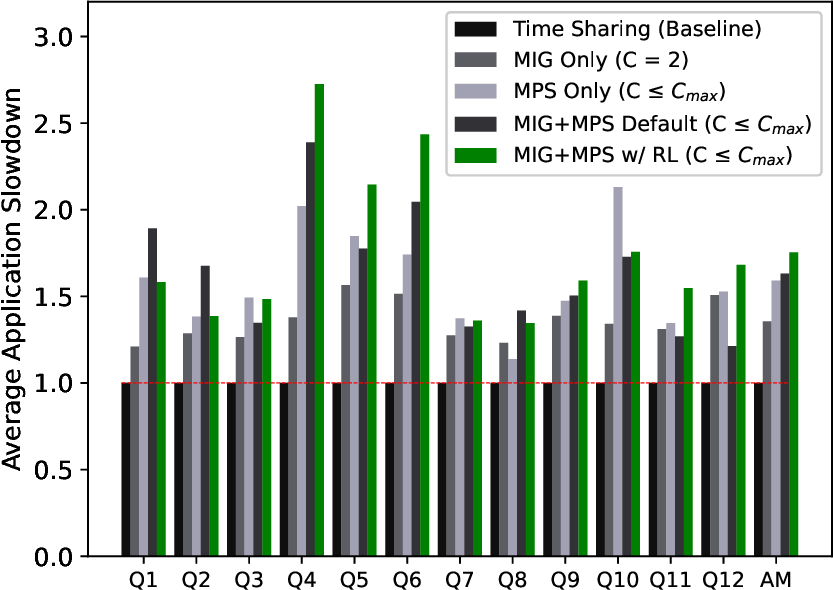}
  \caption{Per Application Slowdown ($C_{max}=4$, $W=12$)}
  \label{fig:slowdown}
\vspace{10pt}
\centering
  \includegraphics[height=5.5cm]{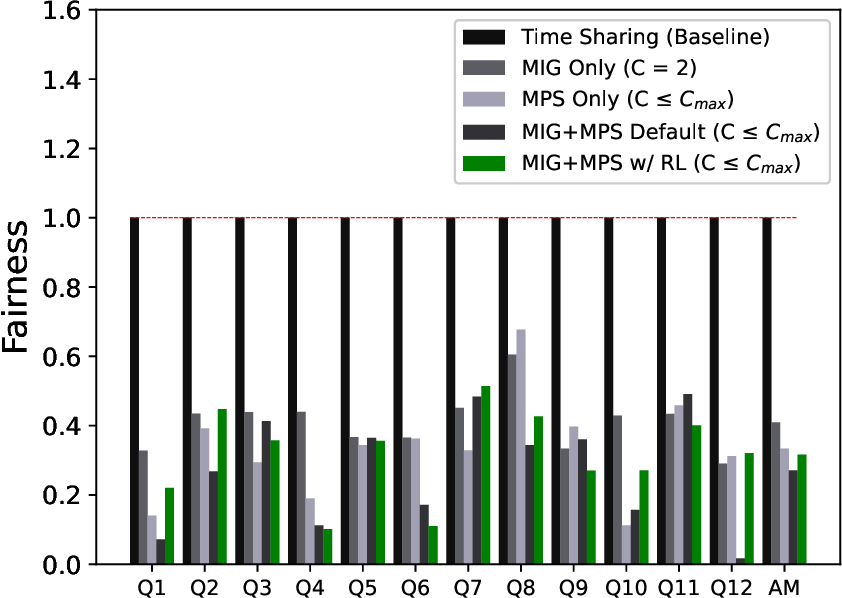}
  \caption{Fairness Comparison ($C_{max}=4$, $W=12$)}
  \label{fig:fairness}
\end{figure}

Next, Figure~\ref{fig:slowdown} demonstrates the average application slowdown caused by co-scheduling for different methods across different job queues. 
The X-axis lists evaluated workloads, while the Y-axis represents the average application slowdown. 
We define the application slowdown ($AppSlowdown$) for a given job taken from the given queue ($J\in\mathrm{Q_i}$) as follows: 
\begin{eqnarray}
AppSlowdown(J)=\frac{CoRunAppTime(J)}{SoloRunAppTime(J)}\nonumber
\end{eqnarray}
Here, $CoRunAppTime(J)$ or $SoloRunAppTime(J)$ denote the space-sharing execution time or the solo-run execution time for the given job ($J$), respectively. 
We calculate the average across all the jobs in the given queue for each method. 

The average application slowdown for our approach is on average 1.829 and is 1.345 at best. 
As co-scheduling can offer more concurrency up to $C_{max}$, it can achieve higher throughput in total as observed in Figure~\ref{fig:tpt} even with the application slowdowns. 
Note that the average application slowdown of \textit{MIG Only ($C=2$)} is smaller than those of the others, however due to the limited concurrency, the total throughout is smaller than the others.
As our approach can trade-off the application slowdowns and concurrency in a better way, it achieves higher total system throughput as a consequence. 

Figure~\ref{fig:fairness} compares the fairness in scheduling among different methods across different workloads. 
By following an existing study~\cite{fairness-metric}, we define the fairness metric ($Fairness$) for the given queue ($\mathrm{Q_i}$) as follows: 
\begin{eqnarray}
Fairness(Q_i)=\frac{\min_{J \in Q_i}(AppSlowdown(J))}{\max_{J \in Q_i}(AppSlowdown(J))}\nonumber
\end{eqnarray}
A higher value is better for this metric, and the highest one is $1$. 
More specifically, when this fairness metric is equal to one, the maximum slowdown becomes exactly the same as the minimum slowdown, which means all the applications suffer from the same degree of slowdown. 
According to Figure~\ref{fig:fairness}, ours is comparable in fairness with the other approaches except for the Time Sharing, even though ours outperforms them in throughput. 
Note we can improve the fairness in our approach by taking it into account in the reward function.




Finally, we report the overhead of our approach in both the online and offline phases. 
The throughput degradation caused by our online optimization is less than 0.5\% on average across our workloads ($W=12$), which is negligible compared with the throughput gain, and thus we observe the considerable throughput improvement, as shown in Figure~\ref{fig:tpt}.
As for the offline training time, a key bottleneck arises due to real-time interactions with the system, i.e., continuous benchmark runs. 
With available MIG/MPS setups for the selected concurrency (let $N_C$ be the number of available setups for $C$, see also TABLE~\ref{part-select}), the maximum count of distinct job selections plus resource assignments is $\sum_{C=2}^{C_{max}} {W \choose C} \times C! \times N_C$. 
Here, to assess the maximum, we suppose selecting $C$ jobs from $W$ unique jobs and assigning them to $C$ distinct regions partitioned with a certain MIG/MPS setup chosen from $N_C$ variants. 
Consequently, for $W=12$ and $C_{max}=4$, the training overhead could escalate to the order of $10^5 \times t_{avg}$, 
where $t_{avg}$ signifies the average duration taken for executing a scheduling policy on the system. However, as the agent progressively converges towards optimal policies, it need not explore every conceivable policy within this set. 
Hence, in our environment, the offline training procedure takes only couple of hours. 
The overhead is reasonable as the training is required only once for a system.

\section{Discussion}\label{discussion}
Our approach is equally extensible to clusters of GPUs because node-local optimizations naturally carry over to clusters and have direct impact on GPU cluster operations. 
To this end, the hierarchical optimization presented in this work needs to be extended by adding another level of resource assignments at the top, i.e., node/GPU allocations. 
For this extension, the vector of job characteristics denoted as $\mathrm{J_i}$ needs to include the numbers of GPUs/nodes requested by the job, which can be retrieved from the corresponding job script. 
Based on this information, the agent will decide the resource allocations denoted as $\mathrm{R_i}$ which also needs to be extended to cover the physical IDs of assigned nodes/GPUs as well as their partitioning states. 
In addition, the agent and the reward function need to coordinately deal with load imbalances introduced by co-scheduling multi-node/-GPU jobs. 
For instance, a multi-node/-GPU job can be co-located with different jobs at different nodes/GPUs which can induce a significant load imbalance for the job.
We consider the following two options for this extension: (1) introducing a larger and more scalable neural network; (2) using a multi-level agent to cope with the system-wide and node-level optimizations separately but coordinately. 



The scenario we are focusing on in this paper are overcrowded systems with long queuing times (i.e., always runnable jobs available). 
This is because they are common in HPC centers with GPU demand going beyond GPU offerings. 
In this situation, we believe it is reasonable and advisable to pick multiple GPU jobs and co-locate them on the same GPU(s) to maximize throughput, and the option for co-starting multiple jobs like our approach can be highly efficient. 
When the system becomes less crowded, a commonly used scheduling policy such as FCFS (First Come First Serve) with backfilling without co-scheduling can be a more efficient option. 
Therefore, in practice, we may choose the policy between them depending on the system state including currently running and queuing jobs. 
Developing such a policy selection mechanism is an interesting research direction and can be one of our future studies in addition to integrating our approach into an existing HPC cluster management tool such as Slurm.

\section{Conclusion and Future Work}\label{Conclusion}
In this paper, we focused on resource partitioning features available in recent commercial GPUs (e.g., MPS and MIG) and proposed a reinforcement learning-based approach to co-optimize the configurations of these multiple and hierarchical resource partitioning features, as well as to make co-scheduling decision for a given set of jobs. 
We observed the impact of that hierarchical resource allocations consisting of MPS and MIG has and based on that defined the matching optimization problem in a concrete mathematical form. We use this formulation to propose our solution based on a reinforcement learning approach.  
Our experimental results showed that our approach was successful in solving the co-optimization problem efficiently.

There are several opportunities to extend this work in the future as discussed in the last section. 
For one, we can extend our work to cover multiple GPUs at the entire cluster scale. 
To this end, the agent and the reward function need to be updated accordingly, by such as using a larger and more scalable neural network or making these entities multi-level, in order for dealing with the increased complexity. 
For implementing this extension, we will consider integrating our approach with an existing HPC cluster management tool such as Slurm. 
Further extensions can include analyzing the impact of application-level resource sharing features (e.g., NVIDIA Multi-Streams~\cite{multi-stream}) on the partitioning features we explored in this paper (MPS and MIG). 
We can consider also other partitioning features on different components as well as other kinds of resources, such as power.



\section*{Acknowledgement}
This work has received funding from the REGALE project from EuroHPC JU under grant agreement no. 956560 and the German Federal Ministry of Education and Research (BMBF) under grant number 16HPC039K. Further, it was supported by BMBF through the initiative SCALEXA and the PDExa project (16ME0641), and by the NVIDIA Academic Hardware Grant Program. Last but not least, we would like to thank professors and staffs in SRD, ITC, The University of Tokyo, in particular Prof. Toshihiro Hanawa, for giving us access to their GPUs.  

{
\renewcommand{\baselinestretch}{0.94}
\bibliographystyle{IEEEtran}
\bibliography{reference}
}

\end{document}